\documentclass[a4paper,11pt]{article}
\usepackage{jheppub}
\usepackage[T1]{fontenc}
\usepackage{amsmath,amssymb,amsfonts}
\usepackage{bm}
\usepackage{graphicx}
\usepackage{hyperref}
\usepackage{orcidlink}
\usepackage{enumitem}
\usepackage{physics}
\usepackage{caption}
\usepackage{tikz}
\usetikzlibrary{decorations.pathmorphing,decorations.markings,arrows.meta}
\usepackage{subcaption}
\usepackage{float}
\usepackage[colorinlistoftodos]{todonotes}

\newcommand{\AdS}{\mathrm{AdS}}
\newcommand{\ESU}{\mathrm{ESU}}
\newcommand{\ii}{\mathrm{i}}

\title{\boldmath Robin boundary conditions in global AdS$_4$: exact double-trace thermodynamics and a soft-mode instability}
\author[a]{David A. Lowe \orcidlink{0000-0002-3168-3612}}
\author[a]{Juanyi Yang \orcidlink{0009-0007-8425-4612}}
\affiliation[a]{Department of Physics, Brown University, Providence, RI, 02912, USA}
\emailAdd{lowe@brown.edu}
\emailAdd{juanyi\_yang@brown.edu}

\abstract{
We consider a conformally coupled scalar field in four-dimensional global anti-de Sitter space with Robin boundary conditions, parametrized by an angle $\alpha$. On the boundary cylinder $\mathbb{R}\times S^{2}$ these conditions realize the double-trace deformation $\tfrac12\lambda\!\int O^{2}$ of the dimension-one operator $O$ in the alternate quantization with $\lambda=\cot\alpha/L$. Because the conformal map to one half of the Einstein static universe is exact, the boundary integral equation can be diagonalized, and the deformed two-point function follows in closed form, $\widehat{\mathcal G}_{\alpha}=\widehat{\mathcal G}_{N}/(1+\lambda\widehat{\mathcal G}_{N})$. Its poles give the normal-mode spectrum, and its determinant gives the free energy exactly within this Gaussian sector. After three local boundary counterterms, the Casimir energy reaches the stability endpoint with a finite square-root cusp. At any finite coupling the bulk $T^{4}$ and $T^{3}$ terms are independent of $\alpha$ and cancel in the difference from Neumann, leaving $\tfrac{\pi}{3}\cot\alpha\,LT^{2}$ as the leading $\alpha$-dependent term. All nonanalyticity comes from one static homogeneous mode, which becomes soft at $\alpha_{\rm crit}$, in agreement with the known classical stability threshold. The susceptibility diverges with exponent $\gamma=1$ and the gap closes with exponent $1/2$. Beyond this angle the mode is tachyonic, and a stable phase would require a stabilizing interaction. In the flat-space limit the physical coupling scales to zero at fixed energy, so the Robin dependence survives only in the soft-frequency sector, which we characterize by a meromorphic Mellin transform in the boost weight. The Robin angle thus gives a control parameter for a Gaussian stability endpoint that can be followed exactly, and raises the analogous question for relaxed boundary conditions in AdS gravity.
}

\begin{document}
\maketitle

\flushbottom

\section{Introduction}
\label{sec:intro}

A scalar field on anti--de Sitter space with mass in the Breitenlohner--Freedman window \cite{Breitenlohner:1982jf,Breitenlohner:1982bm} has two admissible fall-offs at the conformal boundary. Since AdS is not globally hyperbolic, the boundary condition relating them forms part of the definition of the theory and is not fixed by the bulk dynamics \cite{Avis:1977yn,Wald:1980jn,Ishibashi:2003jd,Ishibashi:2004wx}. Setting either fall-off coefficient to zero gives the standard or alternate quantization of the boundary theory. More generally, a linear relation between the coefficients corresponds holographically to a double-trace deformation $\tfrac12\lambda_\Delta\!\int O^2$ of the alternate quantization, where $O$ has dimension $\Delta_-$ and its expectation value is carried by the slow fall-off \cite{Klebanov:1999tb,Witten:2001ua,Berkooz:2002ug}. Inside the open Breitenlohner--Freedman window, $2\Delta_-<d$, so this deformation is relevant. At the Breitenlohner--Freedman bound the two fall-offs degenerate and a logarithmic branch appears. Written instead about the standard quantization, the corresponding double-trace operator has dimension $2\Delta_+>d$ and is irrelevant. The resulting renormalization-group flow and its one-loop realization in the bulk are well understood \cite{Gubser:2002zh,Gubser:2002vv,Hartman:2006dy,Diaz:2007an}.

On the Poincar\'e patch the conformal boundary is flat and introduces no additional length scale. The coupling $\lambda_\Delta$ is therefore the only scale and has positive mass dimension. Its effects grow towards the infrared, so the mixed boundary condition describes a relevant trajectory from the alternate quantization in the ultraviolet to the standard quantization in the infrared. A sufficiently negative coupling destabilizes the theory, and the bulk description develops a tachyonic bound state \cite{Troost:2003mv,Casper:2017pkc}.

The global problem contains an additional scale. The conformal boundary of global AdS$_4$ is the cylinder $\mathbb{R}\times S^2_L$, whose radius provides an infrared scale absent on the Poincar\'e boundary. For the conformally coupled scalar considered here, the Robin coefficient combines with $L$ into the dimensionless parameter $\cot\alpha$. Dirichlet boundary conditions correspond to $\alpha=0$ and Neumann boundary conditions to $\alpha=\pi/2$. At fixed $L$, varying $\alpha$ therefore scans a one-parameter family of generically non-conformal theories rather than following renormalization-group time. Each member is well defined while the corresponding self-adjoint extension of the bulk wave operator remains positive \cite{Ishibashi:2004wx}.

The compact spatial section turns the boundary response into a discrete spectral problem. As $\alpha$ varies, the normal frequencies move continuously. At a critical angle the static homogeneous mode reaches zero frequency, and beyond this point it becomes unstable. The Robin angle therefore controls both the relative weight of the two asymptotic fall-offs and the gap of the global vacuum. Hadamard ground states can be constructed throughout the admissible stable range \cite{Dappiaggi:2016fwc,Dappiaggi:2018pju}. The classical stability threshold and the renormalized vacuum polarization were computed in \cite{Morley:2020ayr}, providing useful benchmarks below \cite{Barroso:2019cwp,Morley:2023mmy}.

Our construction is organized around a single meromorphic response function. The bulk-to-boundary kernel that maps inhomogeneous Robin data on $\mathbb{R}\times S^2$ to the regular bulk solution contains the Robin spectral function in its denominator. Its poles therefore reproduce the complete normal-mode spectrum, while the usual mode expansion follows from its residues. No explicit normalization of the individual modes is required. In Euclidean signature the same kernel is obtained from the boundary restriction of the Robin Green function. Treating the Robin term as a boundary perturbation of the Neumann problem gives the exact boundary integral equation of \cite{Deutsch:1978sc}. After Fourier transformation along $\mathbb{R}$ and harmonic decomposition on $S^2$, the equation becomes algebraic in each channel,
\begin{equation}
\widehat G_\alpha(\omega,\ell)
=\frac{\widehat G_N(\omega,\ell)}
{1+\lambda\,\widehat G_N(\omega,\ell)}\,,
\qquad
\lambda=\frac{\cot\alpha}{L}\,.
\label{eq:resolvent-intro}
\end{equation}
With the spectral normalization $[\widehat G_N]=L$, the denominator is dimensionless. The poles of this resolvent determine the spectrum, its static zero locates the instability, its determinant gives the free energy, and its large-frequency behavior controls the flat-space limit.

Because the bulk scalar is free, the double-trace flow $\partial_\lambda F_\alpha=\tfrac12\langle O^2\rangle_\alpha$ is Gaussian and integrates to a functional determinant \cite{Gubser:2002zh,Hartman:2006dy}. The resulting one-loop expression is exact within the free scalar sector and contains the complete dependence on $\alpha$ at this order. On $S^1_\beta\times S^2$ with $T=1/\beta$,
\begin{equation}
\Delta F(\alpha,T)\equiv F_\alpha-F_N
=\frac{T}{2}\sum_{\ell=0}^{\infty}(2\ell+1)
\sum_{n\in\mathbb Z}
\log\!\left[
1+\lambda\,\widehat G_N^E(\omega_n,\ell)
\right],
\label{eq:detF-intro}
\end{equation}
where the Euclidean Neumann kernel $\widehat G_N^E$ is known in closed form as a ratio of gamma functions.
The determinant separates naturally into vacuum and thermal parts. All ultraviolet divergences reside in the temperature-independent Casimir energy and are removed by three local boundary counterterms. The first scheme-independent coefficient appears at quartic order, $\mathcal E_\alpha^{\rm ren}\simeq-c_4\cot^4\alpha$, with $c_4\approx0.195/L$. As the stability endpoint is approached, the renormalized vacuum energy remains finite but develops a square-root cusp. For any finite Robin coupling, the thermal contribution is ultraviolet finite and has the high-temperature behavior $\tfrac{\pi}{3}\cot\alpha\,LT^2$. The absolute free energy begins with the bulk Stefan--Boltzmann term $-\zeta(4)L^3T^4$, followed by a $T^3$ contribution. For finite Robin coupling these leading terms are the same as at the Neumann point and therefore cancel in $F_\alpha-F_N$ \cite{Allen:1986ty}. The Dirichlet endpoint is nonuniform in this limit and retains a leading $T^3$ difference.

Differentiating the free energy with respect to the coupling gives
\begin{equation}
\langle O^2\rangle_\alpha^{\rm ren}
=
-\frac{2c_4}{\pi L}\cot^3\alpha
+O(\cot^4\alpha).
\end{equation}
The cubic behavior near the Neumann point follows directly from the ultraviolet subtractions and the double-trace expansion. It is not a fitted exponent. The same ultraviolet insensitivity appears in local bulk observables. The renormalized vacuum polarization has an $\alpha$-independent boundary limit for every non-Dirichlet Robin condition \cite{Morley:2020ayr}. In the boundary description, the dependence on $\alpha$ resides in the subtracted infrared part of the determinant, while the leading short-distance behavior remains universal. This is the spectral counterpart of the separation between the universal Hadamard singularity and the regular boundary-condition-dependent terms in \cite{Deutsch:1978sc,Kennedy:1979ar,Kent:2014nya}.

The phase structure is controlled by a single mode. The static factor $1+\lambda\widehat G_N^E(0,\ell)$ is proportional to the inverse susceptibility in the $\ell$-th channel. The homogeneous mode has the largest static response, $\widehat G_N^E(0,0)=\pi L/2$, and is therefore the first to soften. Its gap vanishes at
\begin{equation}
\alpha_{\rm crit}=\pi-\arctan(\pi/2),
\end{equation}
in agreement with the classical stability threshold of \cite{Morley:2020ayr}. Since the static Neumann response contains no temperature dependence, the same value of $\alpha_{\rm crit}$ controls the Gaussian instability at every temperature. The instability line in the $(\alpha,T)$ plane is therefore vertical and meets the zero-temperature axis at a Gaussian critical endpoint.
Near this endpoint, all singular thermodynamic behavior comes from a single oscillator with frequency $\omega_0\propto(\alpha_{\rm crit}-\alpha)^{1/2}$. At zero temperature its zero-point energy produces the square-root cusp. At finite temperature its classical free energy gives the logarithmic divergence, while the heat capacity reduces to a scaling function of $\omega_0/T$. The susceptibility diverges with exponent $\gamma=1$ and the gap closes with exponent $1/2$. These are the Gaussian mean-field exponents associated with the softening of a single mode.
This mechanism is distinct from a Hagedorn transition. A free scalar has only a polynomial density of states, and the nonanalyticity occurs as a function of $\alpha$ rather than temperature. It results from one pole of the exact boundary response reaching zero frequency, not from a collective accumulation of states. The Robin deformation nevertheless shares with $T\bar T$-type deformations the property that the deformed spectrum can be reconstructed from undeformed data through a closed relation \cite{McGough:2016lol,Taylor:2018xcy,Hartman:2018tkw}. The mechanisms are different. $T\bar T$ is irrelevant and reorganizes the ultraviolet spectrum, whereas the present deformation is relevant and leaves the leading ultraviolet behavior unchanged.

The same correlator also provides a useful perspective on the flat-space limit. Near the null-infinity scaling region, the boundary cylinder $\mathbb{R}\times S^2$ connects naturally with the kinematics of celestial holography \cite{Pasterski:2017kqt,Pasterski:2021rjz}. Taking $L\to\infty$ at fixed physical energy sends $\lambda=\cot\alpha/L$ to zero, so fixed-energy two-point functions become independent of $\alpha$. The Robin dependence survives only in the soft-frequency region $w=O(1/L)$, which is also the sector responsible for the instability. We characterize this regime using the Mellin transform of the homogeneous channel. The resulting function is meromorphic in the boost weight, and its infrared residues diverge as $\alpha$ approaches the critical value. Mixed boundary conditions have also been interpreted in terms of double-trace deformations directly in celestial and wedge settings \cite{Fukada:2023vjt}. Here the deformation is instead resummed at finite $L$, where its spectrum and thermodynamics can be followed exactly.

The paper is organized as follows. Section \ref{sec:global_robin} formulates the Robin problem for the conformally coupled scalar in global AdS$_4$, maps it to the half-Einstein static universe, and constructs the bulk-to-boundary kernel whose residue expansion gives the normal-mode sum. Section \ref{sec:euclidean_resolvent} derives the Euclidean boundary resolvent and separates the universal short-distance structure of local observables from the boundary-condition dependence. Section \ref{sec:thermo} develops the thermodynamics, including the functional determinant, its renormalization, the phase structure, the double-trace expectation value, and the soft-mode scaling of the entropy and heat capacity. Section \ref{sec:celestial} studies the flat-space limit and the Mellin-space description of the deformation. Section \ref{sec:conclusions} summarizes the results and discusses extensions.

\section{The global Robin problem and the bulk-to-boundary kernel}
\label{sec:global_robin}

In global coordinates, the $\AdS_{4}$ metric is
\begin{equation}
ds^{2} = \frac{L^{2}}{\cos^{2}\rho}
\left( -dt^{2} +d\rho^{2} +\sin^{2}\rho\, d\Omega_{2}^{2} \right),
\qquad 0\leq\rho<\frac{\pi}{2}\,.
\label{eq:global_metric}
\end{equation}
We work on the covering space, with $t\in\mathbb{R}$. The conformal boundary is the timelike cylinder $\mathbb{R}\times S^{2}$ at $\rho=\pi/2$.

Consider first a scalar of general mass. Near the boundary, a solution of the wave equation behaves as
\begin{equation}
\phi \sim A(t,\Omega)\,(\cos\rho)^{\Delta_{-}}
      - B(t,\Omega)\,(\cos\rho)^{\Delta_{+}},
\qquad
\Delta_{\pm}
=
\frac{3}{2}
\pm
\sqrt{\frac{9}{4}+m^{2}L^{2}}\,.
\label{eq:falloffs}
\end{equation}
For
\begin{equation}
-\frac94<m^2L^2<-\frac54
\end{equation}
both asymptotic behaviors admit the two standard quantizations \cite{Breitenlohner:1982jf,Breitenlohner:1982bm,Klebanov:1999tb}. The endpoints require separate treatment and will not be needed here. Besides the two conditions $A=0$ and $B=0$, one may impose
\begin{equation}
B(t,\Omega)
=
-\,\lambda_{\Delta}L^{\Delta_{+}-\Delta_{-}}\,
A(t,\Omega)\,.
\label{eq:robin_general}
\end{equation}
The explicit power of $L$ is required because the coefficients in \eqref{eq:falloffs} multiply powers of the dimensionless coordinate $\cos\rho$ and therefore carry the same engineering dimension. The coupling $\lambda_\Delta$ has mass dimension $\Delta_+-\Delta_->0$, while
\begin{equation}
g_\Delta
\equiv
\lambda_\Delta L^{\Delta_+-\Delta_-}
\end{equation}
is dimensionless.

In the boundary theory, \eqref{eq:robin_general} describes the double-trace deformation $\tfrac12\lambda_\Delta O^2$ of the alternate quantization \cite{Witten:2001ua,Berkooz:2002ug}. The operator $O$ has dimension $\Delta_-$, its expectation value is carried by the slow fall-off, and the fast fall-off supplies the source after the appropriate normalization by $L$. Since $2\Delta_-<3$ inside the window, the deformation is relevant. Written in terms of the operator of dimension $\Delta_+$ associated with the standard quantization, the corresponding double-trace deformation is irrelevant because $2\Delta_+>3$.
On the Poincar\'e patch the boundary $\mathbb{R}^{1,2}$ is scale-free, so $\lambda_\Delta$ is the only scale and drives a renormalization-group flow between the two quantizations. On global AdS the boundary cylinder carries the radius $L$, and $g_\Delta$ is a fixed dimensionless coupling. Varying the boundary condition at fixed $L$ scans this coupling rather than following renormalization-group time. Each value of $g_\Delta$ therefore defines a distinct, generically non-conformal boundary theory on $\mathbb{R}\times S^2$. The Poincar\'e-patch flow is recovered when the finite-size scale is removed.

We now specialize to the massless conformally coupled scalar, $m^2L^2=-2$, for which
\begin{equation}
(\Delta_-,\Delta_+)=(1,2).
\end{equation}
This case is distinguished in two respects. First, $\Delta_+-\Delta_-=1$. After removing the leading factor of $\cos\rho$, the two asymptotic coefficients become the boundary value and normal derivative of a regular rescaled field. The mixed condition therefore takes the local Robin form
\begin{equation}
\left.
\left(
\cos\alpha\,\widetilde\phi
+
\sin\alpha\,\partial_\rho\widetilde\phi
\right)
\right|_{\rho=\pi/2}
=0\,.
\label{eq:local_robin_homogeneous}
\end{equation}
Second, global AdS$_4$ is conformal to one half of the Einstein static universe and the field is conformally coupled. The transformation
\begin{equation}
\phi=\cos\rho\,\widetilde\phi
\label{eq:conformal_field_relation}
\end{equation}
therefore maps the AdS wave equation and its boundary-value problem exactly to the corresponding problem on the half-ESU.

The conformally related metric is
\begin{equation}
d\widetilde s^{2}
=
L^{2}
\left(
-d t^{2}
+d\rho^{2}
+\sin^{2}\rho\,d\Omega_{2}^{2}
\right),
\qquad
\widetilde g_{\mu\nu}
=
\cos^{2}\rho\,g_{\mu\nu}\,.
\label{eq:esu_metric}
\end{equation}
The rescaled field is smooth at $\rho=\pi/2$. From \eqref{eq:falloffs},
\begin{equation}
\widetilde\phi
=
A-B\cos\rho+\cdots,
\end{equation}
and hence
\begin{equation}
\widetilde\phi|_{\partial}=A,
\qquad
\partial_\rho\widetilde\phi|_{\partial}=B.
\label{eq:falloff_identification}
\end{equation}
This fixes the relative sign in \eqref{eq:falloffs} and the sign convention for the double-trace coupling used below. The rescaled field satisfies
\begin{equation}
\left(
\widetilde\Box-L^{-2}
\right)\widetilde\phi=0\,.
\label{eq:esu_wave}
\end{equation}
The half-Einstein static universe, $0\leq\rho\leq\pi/2$, is a spacetime with a timelike boundary at $\rho=\pi/2$, and the choice of boundary condition forms part of the dynamical problem \cite{Wald:1980jn,Ishibashi:2003jd,Ishibashi:2004wx}.

Expanding in frequency and spherical harmonics,
\begin{equation}
\widetilde\phi(t,\rho,\Omega)
=
e^{-\ii\omega t}
Y_{\ell m}(\Omega)
\psi_{\omega\ell}(\rho),
\label{eq:mode_decomposition}
\end{equation}
gives
\begin{equation}
\left[
\frac{d}{d\rho}
\left(
\sin^{2}\rho
\frac{d}{d\rho}
\right)
-
(1-\omega^{2})\sin^{2}\rho
-
\ell(\ell+1)
\right]
\psi_{\omega\ell}(\rho)
=0\,.
\label{eq:radial_equation}
\end{equation}
The radial equation is independent of the boundary condition. Regularity at $\rho=0$ selects a unique solution up to normalization. A convenient Lorentzian representative is
\begin{equation}
\psi_{\omega\ell}(\rho)
=
(\sin\rho)^{-1/2}
\mathbf Q^{\ell+1/2}_{\omega-1/2}(\cos\rho),
\label{eq:regular_lorentzian}
\end{equation}
where $\mathbf Q^\mu_\nu$ is Olver's normalized Legendre function. In Euclidean signature we use
\begin{equation}
p_{\omega\ell}(\rho)
=
(\sin\rho)^{-1/2}
P^{-\ell-1/2}_{\ii\omega-1/2}(\cos\rho).
\label{eq:regular_euclidean}
\end{equation}
If we include a boundary source, the local Robin condition becomes
\begin{equation}
\left.
\left(
\cos\alpha\,\widetilde\phi
+
\sin\alpha\,\partial_\rho\widetilde\phi
\right)
\right|_{\rho=\pi/2}
=
h(t,\Omega),
\label{eq:inhomogeneous_robin}
\end{equation}
with Dirichlet and Neumann boundary conditions at $\alpha=0$ and $\alpha=\pi/2$, respectively \cite{Dappiaggi:2016fwc,Dappiaggi:2018pju}. Using \eqref{eq:falloff_identification}, one writes
\begin{equation}
B+\cot\alpha\,A
=
\frac{h}{\sin\alpha}.
\end{equation}
Comparison with \eqref{eq:robin_general} gives
\begin{equation}
\lambda = \frac{\cot\alpha}{L},
\label{eq:lambda_definition}
\end{equation}
so that the dimensionless coupling is $\lambda L=\cot\alpha$.
The canonically normalized source for the dimension-one operator is
\begin{equation}
J
=
\frac{h}{L\sin\alpha},
\end{equation}
and the coefficient relation may therefore be written as
\begin{equation}
\frac{B}{L}
+
\lambda A
=
J.
\label{eq:sourced_coefficient_relation}
\end{equation}
This normalization will also be useful below.

\begin{table}[h]
\centering
\begin{tabular}{ll}
\hline
Bulk quantity & Boundary interpretation \\
\hline
$\alpha=\pi/2$ (Neumann) & alternate quantization, $\Delta_O=1$ \\
$\alpha=0$ (Dirichlet) & standard quantization, $\Delta_O=2$ \\
$A$ & $\langle O\rangle$ \\
$B/L$ & source for $O$ \\
$\lambda=\cot\alpha/L$ & coupling of $\tfrac12\lambda O^2$ \\
$\widehat{\mathcal G}_\alpha$ & connected two-point function of $O$ \\
$D_\alpha=0$ & pole of the boundary correlator, bulk normal mode \\
$\tfrac12{\rm Tr}\log(1+\lambda\widehat{\mathcal G}_N)$
& shift of the boundary free energy \\
\hline
\end{tabular}
\caption{Holographic dictionary for the Robin family. The bulk field is free, so the boundary sector is a generalized free field deformed by a double-trace operator.}
\label{tab:dictionary}
\end{table}
The holographic dictionary is summarized in Table \ref{tab:dictionary}, and the same relation follows from a variational principle. With the normalization above, the renormalized bulk action satisfies
\begin{equation}
\delta I_{\rm ren}
=
\int_{\partial}d^3x\sqrt{\gamma}\,
\frac{B}{L}\,\delta A.
\end{equation}
Adding
\begin{equation}
I_{\lambda,J}
=
I_{\rm ren}
+
\int_{\partial}d^3x\sqrt{\gamma}
\left(
\frac12\lambda A^2-JA
\right)
\label{eq:boundary_action}
\end{equation}
gives, upon variation with respect to $A$,
\begin{equation}
\frac{B}{L}
+\lambda A
=
J,
\end{equation}
which is \eqref{eq:sourced_coefficient_relation}. In the Gaussian theory the same variation shifts the inverse two-point function by a constant,
\begin{equation}
\widehat{\mathcal G}_{\lambda}
=
\left(
\widehat{\mathcal G}_N^{-1}
+\lambda
\right)^{-1}
=
\frac{\widehat{\mathcal G}_N}
{1+\lambda\widehat{\mathcal G}_N}.
\label{eq:resolvent_action}
\end{equation}
This relation organizes the analysis below. The spectrum, susceptibility, thermodynamics, instability and flat-space limit all follow from the same denominator.

Not every member of the family is stable. Stability requires the spatial operator defined by the Robin extension to be non-negative \cite{Ishibashi:2004wx}. As the coupling is decreased through negative values, the first loss of positivity occurs in the static homogeneous sector. The lowest $\ell=0$ eigenvalue reaches zero at \cite{Morley:2020ayr}
\begin{equation}
\alpha_{\rm crit}
=
\pi-\arctan\left(\frac{\pi}{2}\right),
\label{eq:critical_alpha}
\end{equation}
after which the corresponding Lorentzian frequency becomes imaginary. We therefore restrict the stable Gaussian theory to
\begin{equation}
0\leq\alpha<\alpha_{\rm crit}.
\end{equation}
At the endpoint the zero mode prevents the existence of a normalizable Gaussian ground state. The approach to this endpoint controls the critical behavior discussed below.

The boundary condition can now be expressed spectrally. Let $\psi_{\omega\ell}$ be the solution of \eqref{eq:radial_equation} regular at the origin. For a source $h_{\omega\ell m}$,
\begin{equation}
\widetilde\phi_{\omega\ell m}(\rho)
=
C_{\omega\ell m}\psi_{\omega\ell}(\rho),
\end{equation}
with
\begin{equation}
C_{\omega\ell m}
=
\frac{h_{\omega\ell m}}
{D_\alpha(\omega,\ell)},
\end{equation}
where
\begin{equation}
D_\alpha(\omega,\ell)
=
\cos\alpha\,
\psi_{\omega\ell}\left(\frac{\pi}{2}\right)
+
\sin\alpha\,
\psi'_{\omega\ell}\left(\frac{\pi}{2}\right).
\label{eq:spectral_function}
\end{equation}
The momentum-space bulk-to-boundary kernel is therefore
\begin{equation}
K_\alpha(\rho;\omega,\ell)
=
\frac{\psi_{\omega\ell}(\rho)}
{D_\alpha(\omega,\ell)}.
\label{eq:kernel_definition}
\end{equation}
A nonzero regular solution of the homogeneous boundary-value problem exists precisely when
\begin{equation}
D_\alpha(\omega,\ell)=0.
\label{eq:normal_mode_condition}
\end{equation}
The normal frequencies $\omega_{n\ell}(\alpha)$ are therefore the poles of $K_\alpha$. Bulk reconstruction and spectral quantization are two representations of the same meromorphic kernel.
At the two conformal endpoints,
\begin{equation}
\omega_{n\ell}
=
\begin{cases}
2n+\ell+1,
& \mathrm{Neumann}\quad(\alpha=\pi/2),
\\[2mm]
2n+\ell+2,
& \mathrm{Dirichlet}\quad(\alpha=0),
\end{cases}
\qquad
n=0,1,2,\ldots,
\end{equation}
in agreement with \cite{Avis:1977yn}. These towers are built on $\Delta_-=1$ and $\Delta_+=2$. Since $\widetilde\phi|_{\partial}=A$ and $\partial_\rho\widetilde\phi|_{\partial}=B$, the condition $\sin\alpha=0$ sets $A=0$ and $\cos\alpha=0$ sets $B=0$. In the homogeneous channel the two sequences may equivalently be identified with the zeros and poles of the analytically continued Neumann response $\tan(\pi\omega/2)/\omega$.

For arbitrary Robin data,
\begin{equation}
\widetilde\phi(t,\rho,\Omega)
=
\sum_{\ell=0}^{\infty}
\sum_{m=-\ell}^{\ell}
\int_{-\infty}^{\infty}
\frac{d\omega}{2\pi}\,
e^{-\ii\omega t}
Y_{\ell m}(\Omega)
K_\alpha(\rho;\omega,\ell)
h_{\omega\ell m}.
\label{eq:bulk_reconstruction}
\end{equation}
Using
\begin{equation}
\sum_{m=-\ell}^{\ell}
Y_{\ell m}(\Omega)
Y_{\ell m}^{*}(\Omega')
=
\frac{2\ell+1}{4\pi}
P_\ell(\cos\gamma),
\label{eq:addition_theorem}
\end{equation}
gives
\begin{equation}
K_\alpha(\rho;\Delta t,\gamma)
=
\sum_{\ell=0}^{\infty}
\frac{2\ell+1}{4\pi}
P_\ell(\cos\gamma)
\int_{-\infty}^{\infty}
\frac{d\omega}{2\pi}\,
e^{-\ii\omega\Delta t}
\frac{\psi_{\omega\ell}(\rho)}
{D_\alpha(\omega,\ell)}.
\label{eq:position_kernel}
\end{equation}
For the retarded solution the poles are displaced according to the causal prescription. Closing the contour in the lower half-plane for $\Delta t>0$ and assuming simple poles gives
\begin{equation}
K_{\alpha}^{\rm ret}(\rho;\Delta t,\gamma)
=
-\ii\Theta(\Delta t)
\sum_{\ell=0}^{\infty}
\frac{2\ell+1}{4\pi}
P_\ell(\cos\gamma)
\sum_n
\frac{
\psi_{\omega_{n\ell},\ell}(\rho)}
{\partial_\omega D_\alpha(\omega_{n\ell},\ell)}
e^{-\ii\omega_{n\ell}\Delta t}.
\label{eq:retarded_mode_sum}
\end{equation}
Restoring the conformal factor,
\begin{equation}
K_{\alpha,\AdS}^{\rm ret}
=
\cos\rho\,
K_\alpha^{\rm ret}.
\end{equation}
Equation \eqref{eq:retarded_mode_sum} is the conventional normal-mode expansion written as a residue sum. The radial wavefunctions arise from the numerator evaluated at the poles, while the spectral weights are determined by derivatives of the Robin spectral function. At $\alpha=\alpha_{\rm crit}$ the lowest pole reaches $\omega=0$. The instability is therefore a single pole of the exact response crossing the origin, while the remaining spectrum stays gapped.

\section{The Euclidean boundary resolvent and local observables}
\label{sec:euclidean_resolvent}

After Wick rotation $t=-\ii\tau$, let $G_\alpha^{\ESU}(x,x')$ satisfy
\begin{equation}
\left(
-\widetilde\Box_E+L^{-2}
\right)
G_\alpha^{\ESU}(x,x')
=
\frac{\delta^{(4)}(x-x')}
{\sqrt{\widetilde g}},
\label{eq:euclidean_green_equation}
\end{equation}
with homogeneous Robin boundary conditions in each argument. Applying Green's identity to a homogeneous solution $\widetilde\phi$ gives
\begin{equation}
\widetilde\phi(x)
=
\int_{\partial V}
\left[
G_\alpha^{\ESU}(x,y)\partial_n\widetilde\phi(y)
-
\widetilde\phi(y)\partial_nG_\alpha^{\ESU}(x,y)
\right]
d S_y.
\label{eq:greens_identity}
\end{equation}
At $\rho=\pi/2$, $\partial_n = L^{-1}\partial_\rho$. The inhomogeneous boundary condition for $\widetilde\phi$ reads
\begin{equation}
\cos\alpha\,\widetilde\phi
+
L\sin\alpha\,\partial_n\widetilde\phi
=
h,
\end{equation}
while the Green function satisfies
\begin{equation}
\cos\alpha\,G_\alpha^{\ESU}
+
L\sin\alpha\,\partial_nG_\alpha^{\ESU}
=
0.
\end{equation}
Substituting these relations into \eqref{eq:greens_identity} cancels the terms proportional to the boundary value of $\widetilde\phi$ and gives
\begin{equation}
\widetilde\phi(x)
=
\frac{1}{L\sin\alpha}
\int_{\mathbb{R}\times S^2}
G_\alpha^{\ESU}(x,y_\partial)
h(y_\partial)
d S_y,
\label{eq:green_reconstruction}
\end{equation}
hence
\begin{equation}
K_\alpha^{\ESU}(x;y_\partial)
=
\frac{1}{L\sin\alpha}
G_\alpha^{\ESU}(x;y_\partial).
\label{eq:kernel_green_relation}
\end{equation}
The bulk-to-boundary kernel is therefore the boundary restriction of the Robin Green function, with the normalization fixed by the Robin boundary operator.

We distinguish four objects in what follows. The quantity $G_\alpha^{\ESU}(x,x')$ is the bulk Green function of the rescaled problem. Its spectral coefficients with both points on the boundary define $\widehat{\mathcal G}_\alpha(\omega,\ell)$, the connected two-point function of $O$. The mixed coefficients $\widehat G_\alpha^{\ESU}(\rho;\omega,\ell)$ retain one radial argument. Finally, $\langle\phi^2\rangle$ is a local bulk observable, whereas $\langle O^2\rangle$ is a renormalized boundary composite operator. The latter is not obtained by simply restricting the former to the boundary.

A useful representation follows by treating the Robin term as a boundary perturbation of the Neumann problem. The Euclidean Green functions obey the exact boundary integral equation \cite{Deutsch:1978sc}
\begin{equation}
G_\alpha^{\ESU}(x,x')
=
G_N^{\ESU}(x,x')
-
\lambda
\int_{\partial V}
G_N^{\ESU}(x,y)
G_\alpha^{\ESU}(y,x')
d S_y,
\qquad
\lambda=\frac{\cot\alpha}{L}.
\label{eq:boundary_dyson}
\end{equation}
This is the boundary analogue of a Dyson equation. The dimensions
\begin{equation}
[G^{\ESU}]=L^{-2},
\qquad
[d S_y]=L^3,
\qquad
[\lambda]=L^{-1}
\end{equation}
make every term in \eqref{eq:boundary_dyson} homogeneous. Translation invariance in Euclidean time and rotational invariance on $S^2$ diagonalize the convolution in the basis $e^{\ii\omega\tau}Y_{\ell m}$.
Define
\begin{equation}
G_N^{\ESU}
\left(
\tau,\rho,\Omega;
\tau',\frac{\pi}{2},\Omega'
\right)
=
\frac{1}{L^3}
\sum_{\ell=0}^{\infty}
\frac{2\ell+1}{4\pi}
P_\ell(\cos\gamma)
\int_{-\infty}^{\infty}
\frac{d\omega}{2\pi}\,
e^{\ii\omega(\tau-\tau')}
\widehat G_N^{\ESU}(\rho;\omega,\ell),
\label{eq:mixed_spectral_decomposition}
\end{equation}
and similarly,
\begin{align}
G_\alpha^{\ESU}
\left(
\tau,\frac{\pi}{2},\Omega;
\tau',\frac{\pi}{2},\Omega'
\right)
=\frac{1}{L^3}
\sum_{\ell=0}^{\infty}
\frac{2\ell+1}{4\pi}
P_\ell(\cos\gamma)\int_{-\infty}^{\infty}
\frac{d\omega}{2\pi}\,
e^{\ii\omega(\tau-\tau')}
\widehat{\mathcal G}_\alpha(\omega,\ell).
\label{eq:boundary_spectral_decomposition}
\end{align}
The factor $L^{-3}$ is chosen so that
\begin{equation}
\left[
\widehat G_N^{\ESU}
\right]
=
\left[
\widehat{\mathcal G}_\alpha
\right]
=
L.
\label{eq:spectral_dimensions}
\end{equation}
Since
\begin{equation}
d S_y
=
L^3d\tau_yd\Omega_y,
\end{equation}
the boundary convolution becomes an ordinary product in each $(\omega,\ell)$ channel:
\begin{equation}
\widehat G_\alpha^{\ESU}(\rho;\omega,\ell)
=
\widehat G_N^{\ESU}(\rho;\omega,\ell)
-
\lambda
\widehat G_N^{\ESU}(\rho;\omega,\ell)
\widehat{\mathcal G}_\alpha(\omega,\ell).
\label{eq:mixed_resolvent_equation}
\end{equation}
Restricting the remaining bulk point to the boundary gives
\begin{equation}
\widehat{\mathcal G}_\alpha(\omega,\ell)
=
\widehat{\mathcal G}_N(\omega,\ell)
-
\lambda
\widehat{\mathcal G}_N(\omega,\ell)
\widehat{\mathcal G}_\alpha(\omega,\ell),
\label{eq:boundary_resolvent_equation}
\end{equation}
and therefore
\begin{equation}
\widehat{\mathcal G}_\alpha(\omega,\ell)
=
\frac{
\widehat{\mathcal G}_N(\omega,\ell)}
{1+\lambda\widehat{\mathcal G}_N(\omega,\ell)}.
\label{eq:dressed_boundary_propagator}
\end{equation}
The denominator is dimensionless because
\begin{equation}
[\lambda\widehat{\mathcal G}_N]=1.
\end{equation}
Likewise,
\begin{equation}
\widehat G_\alpha^{\ESU}(\rho;\omega,\ell)
=
\frac{
\widehat G_N^{\ESU}(\rho;\omega,\ell)}
{1+\lambda\widehat{\mathcal G}_N(\omega,\ell)},
\label{eq:dressed_mixed_propagator}
\end{equation}
and combine this with \eqref{eq:kernel_green_relation} gives
\begin{align}
\widehat K_\alpha^{\ESU}(\rho;\omega,\ell)
&=
\frac{1}{L\sin\alpha}
\frac{
\widehat G_N^{\ESU}(\rho;\omega,\ell)}
{1+\dfrac{\cot\alpha}{L}
\widehat{\mathcal G}_N(\omega,\ell)}
\nonumber\\
&=
\frac{
\widehat G_N^{\ESU}(\rho;\omega,\ell)}
{L\sin\alpha+\cos\alpha\,
\widehat{\mathcal G}_N(\omega,\ell)}.
\label{eq:exact_spectral_kernel}
\end{align}
Equations \eqref{eq:dressed_boundary_propagator} and \eqref{eq:exact_spectral_kernel} are exact, and no expansion in the Robin parameter is involved, and the radial quantization condition need not be solved mode by mode. The deformation is encoded entirely in one scalar denominator in each channel \cite{Klebanov:1999tb,Witten:2001ua,Hartman:2006dy,Wang:2026esp}.

After analytic continuation, the Lorentzian normal frequencies satisfy
\begin{equation}
1+\lambda
\widehat{\mathcal G}_N^L(\omega,\ell)
=
0,
\label{eq:resolvent_pole_condition}
\end{equation}
which is equivalent to $D_\alpha(\omega,\ell)=0$. The radial and Green-function constructions thus give the same spectral condition. The first zero at $\omega=0$ occurs in the homogeneous channel,
\begin{equation}
1+\lambda
\widehat{\mathcal G}_N(0,0)
=
0.
\label{eq:critical_resolvent}
\end{equation}
For the conformal scalar,
\begin{equation}
\widehat{\mathcal G}_N(0,0)
=
\frac{\pi L}{2},
\end{equation}
so
\begin{equation}
1+\frac{\pi}{2}
\cot\alpha_{\rm crit}
=
0,
\end{equation}
or
\begin{equation}
\alpha_{\rm crit}
=
\pi-\arctan\left(\frac{\pi}{2}\right).
\end{equation}
This agrees with the classical stability threshold of \cite{Morley:2020ayr}. The appearance of an imaginary-frequency bulk mode for negative double-trace coupling is the global counterpart of the Poincar\'e-patch instability discussed in \cite{Troost:2003mv,Casper:2017pkc}.

Near the endpoint,
\begin{equation}
1+\lambda
\widehat{\mathcal G}_N^L(\omega,0)
=
\varepsilon(\alpha)
-
\kappa\omega^2
+
O(\omega^4),
\label{eq:soft_expansion}
\end{equation}
where
\begin{equation}
\varepsilon(\alpha_{\rm crit})=0,
\qquad
\kappa>0.
\end{equation}
The lowest frequency therefore satisfies
\begin{equation}
\omega_0^2
\simeq
\frac{\varepsilon(\alpha)}{\kappa}.
\label{eq:gap_closing}
\end{equation}
Since $\varepsilon$ vanishes linearly in $\alpha_{\rm crit}-\alpha$, the frequency itself closes as the square root of the distance from the critical boundary condition.

The Euclidean formulation also separates the universal short-distance singularity from the boundary-condition dependence. For the conformal scalar, the Dirichlet and Neumann AdS Green functions have the image representations \cite{Avis:1977yn,Deutsch:1978sc}
\begin{align}
G_D^E(x,x')
&=
\frac{\cos\rho\cos\rho'}{8\pi^2L^2}
\left[
\frac{1}{\cosh\Delta\tau-\cos\Psi}
-
\frac{1}{\cosh\Delta\tau-\cos\Psi'}
\right],
\label{eq:dirichlet_image}
\\
G_N^E(x,x')
&=
\frac{\cos\rho\cos\rho'}{8\pi^2L^2}
\left[
\frac{1}{\cosh\Delta\tau-\cos\Psi}
+
\frac{1}{\cosh\Delta\tau-\cos\Psi'}
\right],
\label{eq:neumann_image}
\end{align}
where
\begin{equation}
\cos\Psi
=
\cos\rho\cos\rho'
+
\sin\rho\sin\rho'\cos\gamma,
\qquad
\cos\Psi'
=
-\cos\rho\cos\rho'
+
\sin\rho\sin\rho'\cos\gamma.
\label{eq:geodesic_separations}
\end{equation}
The first term contains the coincident Hadamard singularity \cite{Kent:2014nya}. The image term is regular at coincidence for fixed interior points and carries the dependence on the boundary condition.

Conformally transforming \eqref{eq:boundary_dyson} back to AdS$_4$ gives
\begin{equation}
G_\alpha^E(x,x')
=
G_N^E(x,x')
-
\lambda\cos\rho\cos\rho'
\int_{\partial V}
G_N^{\ESU}(x,y)
G_\alpha^{\ESU}(y,x')
d S_y.
\label{eq:ads_boundary_integral}
\end{equation}
The second term is finite in the coincidence limit. Hence
\begin{align}
\left\langle\phi^2(x)\right\rangle_\alpha^{\rm ren}
-
\left\langle\phi^2(x)\right\rangle_N^{\rm ren}
=
-\lambda\cos^2\rho
\int_{\partial V}
G_N^{\ESU}(x,y)
G_\alpha^{\ESU}(y,x)
d S_y.
\label{eq:vacuum_polarization_difference}
\end{align}
No additional coincident-point subtraction is needed on the right-hand side.

Formally iterating the integral equation gives
\begin{equation}
G_\alpha^{\ESU}
=
G_N^{\ESU}
-
\lambda G_N^{\ESU}\star G_N^{\ESU}
+
\lambda^2
G_N^{\ESU}\star G_N^{\ESU}\star G_N^{\ESU}
-\cdots,
\label{eq:boundary_neumann_series}
\end{equation}
where $\star$ denotes convolution over $\mathbb{R}\times S^2$. This expansion is useful for identifying the near-boundary scaling, although \eqref{eq:dressed_boundary_propagator} gives its exact resummation.

Let
\begin{equation}
\epsilon
=
\frac{\pi}{2}-\rho.
\end{equation}
The boundary-induced part of the Neumann coincidence function on the ESU behaves as $\epsilon^{-2}$. The first boundary insertion behaves as $\epsilon^{-1}$, while subsequent insertions remain finite. Since
\begin{equation}
\cos^2\rho\sim\epsilon^2,
\end{equation}
all $\alpha$-dependent corrections vanish after the conformal transformation in the boundary limit. Therefore
\begin{equation}
\lim_{\rho\to\pi/2}
\left\langle\phi^2(x)\right\rangle_\alpha^{\rm ren}
=
\lim_{\rho\to\pi/2}
\left\langle\phi^2(x)\right\rangle_N^{\rm ren},
\qquad
0<\alpha<\alpha_{\rm crit}.
\label{eq:universal_boundary_limit}
\end{equation}
For the conformal vacuum \cite{Morley:2020ayr},
\begin{equation}
\lim_{\rho\to\pi/2}
\left\langle\phi^2(x)\right\rangle_\alpha^{\rm ren}
=
\frac{5}{48\pi^2L^2},
\qquad
\alpha\neq0.
\label{eq:boundary_value}
\end{equation}
The Dirichlet endpoint is exceptional because it fixes the boundary value of the rescaled field itself. For every non-Dirichlet Robin condition the leading boundary behavior is Neumann-like. The boundary parameter changes the global response and the low-lying spectrum but not the leading local boundary limit.

The Robin family therefore separates two effects of the boundary condition. Soft and global observables retain a strong dependence on $\alpha$ through the positions of the poles and the lowest gap. Local near-boundary observables are controlled by the universal short-distance structure and lose this dependence at leading order. The same distinction should organize the renormalized stress-energy tensor computed for this family in \cite{Morley:2023mmy}.

\section{Thermodynamics of the Robin family}
\label{sec:thermo}

The algebraic dressing \eqref{eq:resolvent_action} is sufficient to determine the free energy of the Gaussian boundary sector. We integrate the double-trace flow and study its dependence on the Robin angle and temperature. Throughout this section the mode sums and frequency integrals are evaluated in units $L=1$. Dimensionful results are restored explicitly where needed.

Turning on the Robin coupling deforms the Neumann theory according to
\begin{equation}
S_\alpha
=
S_N
+
\frac{\lambda}{2}
\int_{\partial}d^3x\sqrt{\gamma}\,O^2,
\qquad
\lambda=\frac{\cot\alpha}{L}.
\label{eq:lambda}
\end{equation}
The deformation is quadratic, so
\begin{equation}
\widehat G_\alpha^{-1}
=
\widehat G_N^{-1}
+\lambda,
\end{equation}
or
\begin{equation}
\widehat G_\alpha(\omega,\ell)
=
\frac{\widehat G_N(\omega,\ell)}
{1+\lambda\widehat G_N(\omega,\ell)}.
\label{eq:Galpha}
\end{equation}
Integrating $\partial_\lambda F_\alpha=\tfrac12\langle O^2\rangle_\alpha$ gives \cite{Gubser:2002zh,Hartman:2006dy,Diaz:2007an}
\begin{equation}
-\log\frac{Z_\alpha}{Z_N}
=
\frac12
{\rm Tr}\log
\left(
1+\lambda\widehat G_N
\right).
\end{equation}
On $S^1_\beta\times S^2$, with bosonic Matsubara frequencies $\omega_n=2\pi nT$ in the $L=1$ convention,
\begin{equation}
\Delta F(\alpha,T)
\equiv
F_\alpha-F_N
=
\frac{T}{2}
\sum_{\ell=0}^{\infty}(2\ell+1)
\sum_{n\in\mathbb Z}
\log
\left[
1+\lambda
\widehat G_N^E(\omega_n,\ell)
\right].
\label{eq:DeltaF}
\end{equation}
The Euclidean kernel $\widehat G_N^E(\Omega,\ell)\equiv\widehat G_N(i\Omega,\ell)$ is real, positive and free of the Lorentzian normal-mode poles.

As $T\to0$, the Matsubara spacing vanishes and the sum becomes an integral,
\begin{equation}
\mathcal E_\alpha
=
\frac12
\sum_{\ell=0}^{\infty}(2\ell+1)
\int_{-\infty}^{\infty}
\frac{d\Omega}{2\pi}
\log
\left[
1+\lambda
\widehat G_N^E(\Omega,\ell)
\right]
=
E_0^\alpha-E_0^N,
\label{eq:Evac}
\end{equation}
where energies are measured in units $1/L$ in this formula. We write
\begin{equation}
\Delta F^{\rm ren}(\alpha,T)
=
\mathcal E_\alpha^{\rm ren}
+
F_{\rm th}(\alpha,T),
\qquad
F_{\rm th}
\equiv
\Delta F-\mathcal E_\alpha.
\label{eq:separate}
\end{equation}
The finite radius $L$ means that varying $\alpha$ at fixed geometry moves through a line of non-conformal theories rather than along renormalization-group time.

The calculation depends on the Euclidean Neumann response
\begin{equation}
\widehat G_N^E(\Omega,\ell)
=
\frac{L}{2}
\frac{
\left|
\Gamma\!\left(
\frac{\ell+1+\ii\Omega}{2}
\right)
\right|^2}
{
\left|
\Gamma\!\left(
\frac{\ell+2+\ii\Omega}{2}
\right)
\right|^2},
\label{eq:GEN}
\end{equation}
where $\Omega$ is dimensionless. It is smooth and positive for real $\Omega$. The first two channels are
\begin{equation}
\widehat G_N^E(\Omega,0)
=
\frac{L}{\Omega}
\tanh\frac{\pi\Omega}{2},
\qquad
\widehat G_N^E(\Omega,1)
=
\frac{L\Omega}{1+\Omega^2}
\coth\frac{\pi\Omega}{2}.
\label{eq:GEN01}
\end{equation}
The remaining channels follow from
\begin{equation}
\frac{
\widehat G_N^E(\Omega,\ell+2)}
{\widehat G_N^E(\Omega,\ell)}
=
\frac{
(\ell+1)^2+\Omega^2}
{
(\ell+2)^2+\Omega^2},
\label{eq:recursion}
\end{equation}
which follows directly from $\Gamma(z+1)=z\Gamma(z)$. Thus even $\ell$ contain a $\tanh(\pi\Omega/2)$ factor and odd $\ell$ a $\coth(\pi\Omega/2)$ factor, multiplied by rational functions of $\Omega^2$.

At large frequency and angular momentum,
\begin{equation}
\widehat G_N^E
\sim
\frac{L}{
\sqrt{\Omega^2+(\ell+1)^2}},
\end{equation}
which is the flat-space ultraviolet behavior of a dimension-one operator. The static values begin as
\begin{equation}
\widehat G_N^E(0,\ell)
=
\frac{\pi L}{2},
\quad
\frac{2L}{\pi},
\quad
\frac{\pi L}{8},
\ldots,
\label{eq:static}
\end{equation}
and decrease with $\ell$. The homogeneous channel therefore has the largest static response.

\subsection{Renormalization and the Casimir energy}
\label{sec:vacuum}

The asymptotic behavior $\widehat G_N^E\sim L/\sqrt{\Omega^2+\ell^2}$ makes the bare determinant \eqref{eq:Evac} ultraviolet divergent. Power counting leaves three divergent orders. The term linear in $\lambda$ is quadratically divergent, the quadratic term is linearly divergent, and the cubic term is logarithmically divergent. All three are temperature independent and correspond to local boundary counterterms \cite{Kennedy:1979ar}.
The thermal difference $F_{\rm th}$ is therefore ultraviolet finite. The ultraviolet terms are identical in the Matsubara sum and its zero-temperature continuum limit and cancel in their difference. Equivalently, after Poisson resummation the thermal part is built from nonzero thermal images. The remaining angular-momentum sum is exponentially suppressed at large $\ell$.

We choose the renormalization prescription
\begin{equation}
\mathcal E_\alpha^{\rm ren}
=
\frac12
\sum_{\ell}(2\ell+1)
\int
\frac{d\Omega}{2\pi}
\left[
\log(1+u)
-u
+\frac{u^2}{2}
-\frac{u^3}{3}
\right],
\qquad
u=\lambda\widehat G_N^E(\Omega,\ell),
\label{eq:Eren}
\end{equation}
where the integrand begins at order $u^4$. Finite local counterterms still permit shifts by
\begin{equation}
c_1\cot\alpha
+
c_2\cot^2\alpha
+
c_3\cot^3\alpha,
\end{equation}
so the absolute value of $\mathcal E_\alpha^{\rm ren}$ remains scheme dependent through cubic order \cite{Gubser:2002vv}. Coefficients beginning at quartic order, together with the nonanalytic part near the stability endpoint, are scheme independent.

The Dirichlet limit $\alpha\to0$, $\lambda\to\infty$ is a nonuniform strong-coupling limit in this parametrization. There the $\Delta_+$ tower takes over, and the weak-coupling expansion about Neumann is no longer appropriate. The subtracted expression \eqref{eq:Eren} is real throughout the stable branch, where
\begin{equation}
1+\frac{\pi}{2}\cot\alpha>0.
\end{equation}

Near the Neumann point,
\begin{equation}
\mathcal E_\alpha^{\rm ren}
=
-c_4\cot^4\alpha
+
O(\cot^5\alpha),
\label{eq:c4onset}
\end{equation}
with
\begin{equation}
c_4
=
\frac{1}{8L}
\sum_{\ell}(2\ell+1)
\int
\frac{d\Omega}{2\pi}
\left(
\frac{\widehat G_N^E}{L}
\right)^4
\simeq
\frac{0.195}{L}.
\label{eq:c4}
\end{equation}
The homogeneous channel contributes approximately $0.139/L$ to this coefficient, and the large-$\ell$ summand falls as $1/(16\ell^2)$, so the coefficient is convergent. Direct integration gives the same result as the $\cot\alpha\to0$ limit of $\mathcal E_\alpha^{\rm ren}/\cot^4\alpha$, and the next coefficient is $c_5L\simeq0.166$.

For $\alpha>\pi/2$, the curve develops a square-root cusp as $\alpha\to\alpha_{\rm crit}$. Define
\begin{equation}
\varepsilon(\alpha)
=
1+\frac{\pi}{2}\cot\alpha,
\end{equation}
and use
\begin{equation}
\widehat G_N^E(\Omega,0)
=
\frac{\pi}{2}
-
\frac{\pi^3}{24}\Omega^2
+\cdots,
\end{equation}
the soft part of the logarithm is
\begin{equation}
\log
\left(
\varepsilon+\kappa\Omega^2
\right),
\qquad
\kappa\rightarrow\frac{\pi^2}{12}.
\end{equation}
Near the endpoint,
\begin{equation}
\varepsilon
=
c_\varepsilon
(\alpha_{\rm crit}-\alpha)
+\cdots,
\qquad
c_\varepsilon
=
\frac{\pi}{2}
+\frac{2}{\pi}.
\end{equation}
The singular part of the integral is proportional to $\sqrt{\varepsilon/\kappa}$, giving
\begin{equation}
\mathcal E_\alpha^{\rm ren}
\simeq
\mathcal E_{\rm crit}^{\rm ren}
+
B\sqrt{\alpha_{\rm crit}-\alpha},
\qquad
B
=
\frac12
\sqrt{\frac{c_\varepsilon}{\kappa}}
\simeq
0.819.
\label{eq:cusp}
\end{equation}
A direct numerical fit gives $B=0.820$. The endpoint value $\mathcal E_{\rm crit}^{\rm ren}\simeq-0.18/L$ refers to the subtraction scheme \eqref{eq:Eren} and is not itself universal. By contrast, the square-root exponent and the coefficient $B$ are unaffected by finite local counterterms. The cusp is the zero-point energy of the mode that becomes soft.

\subsection{The phase diagram and the soft mode}
\label{sec:phase}

The thermal term is evaluated as the Matsubara sum minus its zero-temperature integral, channel by channel. The angular-momentum sum converges exponentially. The result vanishes at Neumann, is positive for $\alpha<\pi/2$, and is negative for $\pi/2<\alpha<\alpha_{\rm crit}$. For any finite Robin coupling its leading high-temperature behavior is proportional to $T^2$.

The power follows from dimensional analysis. At leading order,
\begin{equation}
\Delta F
\sim\frac{\lambda}{2}
\int_{S^2}\sqrt{\gamma}\,
\langle O^2\rangle_T,
\end{equation}
and $O^2$ has dimension two in the ultraviolet Neumann theory. Hence $\langle O^2\rangle_T\propto T^2$. Linearizing the logarithm and replacing the Neumann kernel by its ultraviolet form gives
\begin{equation}
F_{\rm th}(\alpha,T)
\to
\frac{\pi}{3}\lambda L^2T^2
=
\frac{\pi}{3}\cot\alpha\,LT^2,
\qquad
T\gg L^{-1}.
\label{eq:highT}
\end{equation}
Equivalently,
\begin{equation}
\langle O^2\rangle_T
\to
\frac{T^2}{6}
\end{equation}
in the normalization in which $\lambda$ couples to $\tfrac12\int\sqrt{\gamma}\,O^2$.

The exact Matsubara sum approaches \eqref{eq:highT} with subleading $O((LT)^{-1})$ corrections. At $\lambda L=-10^{-2}$ we find
\begin{equation}
\frac{F_{\rm th}}{\lambda L^2T^2}
=
1.0382,\quad
1.0429,\quad
1.0444
\end{equation}
at $LT=32,64,96$, respectively, approaching $\pi/3=1.0472$.

The origin of the high-temperature powers is clearest at the conformal endpoints. Consider their one-particle partition functions are
\begin{equation}
z_\Delta(q)
=
\sum_{n,\ell}
(2\ell+1)
q^{2n+\ell+\Delta}
=
\frac{q^\Delta}{(1-q)^3},
\qquad
q=e^{-\beta/L}.
\label{eq:oneparticle}
\end{equation}
The corresponding high-temperature expansion is
\begin{equation}
F_\Delta
=
-\zeta(4)L^3T^4
-
\left(
\frac32-\Delta
\right)
\zeta(3)L^2T^3
+\cdots,
\label{eq:absoluteF}
\end{equation}
and the $T^4$ term is the Stefan--Boltzmann contribution from three bulk spatial dimensions \cite{Allen:1986ty}. For every finite Robin coupling, the ultraviolet spectrum approaches the Neumann spectrum, so the $T^4$ and $T^3$ terms are independent of $\alpha$ and cancel in $F_\alpha-F_N$. Equation \eqref{eq:highT} is then the first surviving $\alpha$-dependent term.
The Dirichlet endpoint is exceptional because it is reached only at $\lambda\to+\infty$. Its ultraviolet spectrum belongs to the $\Delta_+=2$ quantization, and therefore
\begin{equation}
F_D-F_N
\sim
\zeta(3)L^2T^3
\qquad
(T\to\infty),
\end{equation}
rather than \eqref{eq:highT}. This is an asymptotic statement, not an exact identity at finite temperature.

The static factor
\begin{equation}
1+\lambda\widehat G_N^E(0,\ell)
=
\frac{
\widehat G_N^E(0,\ell)}
{\widehat G_\alpha^E(0,\ell)}
\propto
m_{\rm eff}^2(\ell)
\label{eq:gap}
\end{equation}
is proportional to the inverse static susceptibility. Since the homogeneous mode has the largest static Neumann response, it is the first to soften. Its inverse susceptibility vanishes at
\begin{equation}
\alpha_{\rm crit}
=
\pi-\arctan\frac{\pi}{2}
\simeq
0.6805\,\pi,
\qquad
\lambda_{\rm crit}
=
-\frac{2}{\pi L}.
\label{eq:acrit}
\end{equation}
Only the $\ell=0$ channel reaches zero at this point; all $\ell\geq1$ channels remain gapped. Beyond $\alpha_{\rm crit}$ the homogeneous mode is tachyonic and the Gaussian vacuum is unstable \cite{Troost:2003mv,Casper:2017pkc}. The free theory therefore establishes a stability endpoint rather than a stable symmetry-broken phase.

Near the endpoint,
\begin{equation}
\widehat G_\alpha^E(\Omega,0)
\simeq
\frac{\pi/2}
{\varepsilon+\kappa\Omega^2}
=
\frac{\pi/2}{\kappa}
\frac{1}{\Omega^2+\omega_0^2},
\qquad
\omega_0
=
2B
\sqrt{\alpha_{\rm crit}-\alpha}.
\label{eq:softpole}
\end{equation}
Thus the frequency squared vanishes linearly in $\alpha_{\rm crit}-\alpha$, while the gap itself closes with exponent $1/2$.
The singular contribution to the free energy is that of one harmonic oscillator,
\begin{equation}
F^{\rm soft}(\alpha,T)
=
\frac{\omega_0}{2}
+
T
\log
\left(
1-e^{-\omega_0/T}
\right).
\label{eq:Fsoft}
\end{equation}
At $T=0$ this reduces to
\begin{equation}
\frac{\omega_0}{2}
=
B\sqrt{\alpha_{\rm crit}-\alpha},
\end{equation}
which reproduces \eqref{eq:cusp}. At fixed $T>0$ with $\omega_0\ll T$,
\begin{equation}
F^{\rm soft}
=
T\log\frac{\omega_0}{T}
+\cdots
\simeq
\frac{T}{2}
\log(\alpha_{\rm crit}-\alpha)
+\cdots,
\end{equation}
and therefore diverges to $-\infty$. The numerical coefficient of the logarithm approaches the single-mode value $T/2$.
The static susceptibility is
\begin{equation}
\chi_0
=
\widehat G_\alpha^E(0,0)
=
\frac{\pi L/2}
{1+\tfrac{\pi}{2}\cot\alpha}
\propto
(\alpha_{\rm crit}-\alpha)^{-1},
\label{eq:chi0}
\end{equation}
and thus we obtain $\chi_0\propto\omega_0^{-2}$. The susceptibility exponent is $\gamma=1$, while the gap exponent is $1/2$. These are Gaussian mean-field exponents, and they describe the spectral softening of one mode and do not by themselves establish a thermodynamic universality class. Since the spatial sphere is compact, only one degree of freedom becomes soft. At the endpoint the corresponding Gaussian zero-mode integral is non-normalizable.

The threshold is independent of $T$ because the critical mode is the static Matsubara mode, $n=0$, in the homogeneous angular channel, $\ell=0$. The free-theory instability line is therefore vertical in the $(\alpha,T)$ plane and meets the $T=0$ axis at $\alpha_{\rm crit}$. The distinction between $T=0$ and $T>0$ is important. At strictly zero temperature an isolated point at $\Omega=0$ has zero measure in \eqref{eq:Evac}, leaving only the finite square-root cusp. At any nonzero temperature the discrete $n=0$ term has finite weight and drives the free energy to $-\infty$.

Note that this behavior is not Hagedorn-like since a Hagedorn singularity is produced by an exponentially growing density of states and occurs at a critical temperature. Here the density of states is polynomial. The nonanalyticity instead occurs as a function of the Robin coupling and is generated by one mode becoming soft. No collective accumulation of states is involved.

\begin{figure}[t]
\centering
\includegraphics[width=\textwidth]{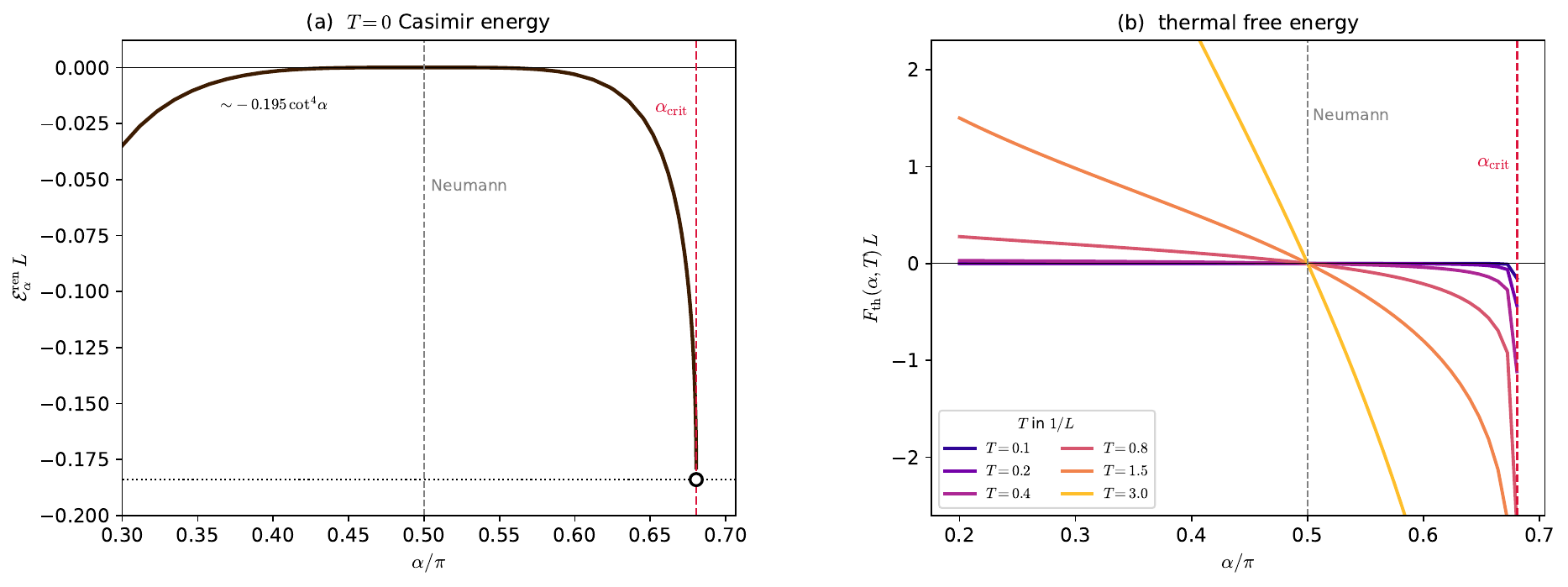}
\caption{The deformed free energy of the Robin family ($L=1$). \textbf{(a)} The renormalized zero-temperature Casimir energy $\mathcal E_\alpha^{\rm ren}$ of \eqref{eq:Eren}. It vanishes at Neumann, begins as $-c_4\cot^4\alpha$ with $c_4\simeq0.195$, and approaches $\alpha_{\rm crit}$ with the finite square-root cusp \eqref{eq:cusp}. The open circle marks the scheme-dependent limiting value $\mathcal E_{\rm crit}^{\rm ren}\simeq-0.18$. \textbf{(b)} The ultraviolet-finite thermal contribution $F_{\rm th}(\alpha,T)$. For finite Robin coupling it grows as $T^2$ at high temperature and diverges to $-\infty$ as $\alpha\to\alpha_{\rm crit}$ at any $T>0$.}
\label{fig:freeenergy}
\end{figure}

\begin{figure}[t]
\centering
\includegraphics[width=\textwidth]{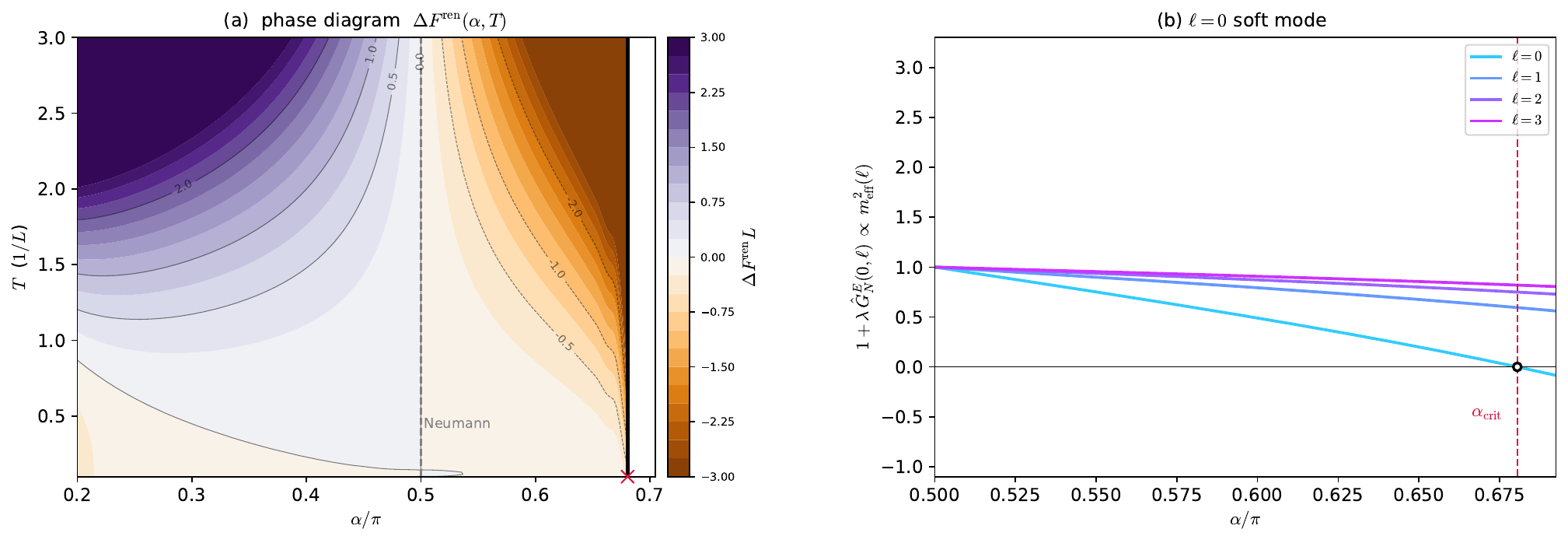}
\caption{Gaussian stability structure of the Robin family ($L=1$). \textbf{(a)} The renormalized free energy $\Delta F^{\rm ren}=\mathcal E_\alpha^{\rm ren}+F_{\rm th}$ in the $(\alpha,T)$ plane. The Gaussian vacuum is stable for $0\leq\alpha<\alpha_{\rm crit}$ and unstable beyond. The instability line is vertical because it is fixed by the temperature-independent static mode. \textbf{(b)} The static stability function $1+\lambda\widehat G_N^E(0,\ell)\propto m_{\rm eff}^2(\ell)$. Only the homogeneous channel reaches zero at $\alpha_{\rm crit}=\pi-\arctan(\pi/2)$; every $\ell\geq1$ channel remains gapped.}
\label{fig:phase}
\end{figure}

\subsection{The double-trace expectation value $\langle O^2\rangle_\alpha$ and ultraviolet universality}
\label{sec:doubletrace}

The free energy is a generating functional for the operator that drives the deformation. Since $\lambda$ couples to $\tfrac12\int_{S^2}\sqrt{\gamma}\,O^2$,
\begin{equation}
4\pi L^2
\langle O^2\rangle_\alpha^{\rm ren}
=
2\partial_\lambda
\mathcal E_\alpha^{\rm ren}.
\end{equation}
Restoring the factors of $L$ in the $L=1$ mode integral gives
\begin{equation}
4\pi L^2
\langle O^2\rangle_\alpha^{\rm ren}
=
-\frac{1}{L}
\sum_{\ell=0}^{\infty}(2\ell+1)
\int_{-\infty}^{\infty}
\frac{d\Omega}{2\pi}
\frac{
\lambda^3
\left[
\widehat G_N^E(\Omega,\ell)
\right]^4}
{
1+\lambda\widehat G_N^E(\Omega,\ell)}.
\label{eq:O2def}
\end{equation}
The same three subtractions as in \eqref{eq:Eren} are understood. Indeed,
\begin{equation}
\frac{d}{d u}
\left[
\log(1+u)-u+\frac{u^2}{2}-\frac{u^3}{3}
\right]
=
-\frac{u^3}{1+u}.
\end{equation}
The compact expression \eqref{eq:O2def} is therefore exact in the Gaussian theory.

Throughout the stable branch,
\begin{equation}
\langle O\rangle=0.
\end{equation}
Equation \eqref{eq:O2def} measures fluctuations of $O$ and the response of the vacuum energy to the double-trace coupling. It is not an order parameter for a broken phase. In the free theory no stable equilibrium state exists beyond $\alpha_{\rm crit}$. If an interaction stabilized that regime, $\langle O\rangle$ rather than $\langle O^2\rangle$ would provide the corresponding order parameter.

Near Neumann,
\begin{equation}
4\pi L^2
\langle O^2\rangle_\alpha^{\rm ren}
=
-8c_4L\,\cot^3\alpha
+
O(\cot^4\alpha),
\qquad
8c_4L\simeq1.557.
\label{eq:O2near}
\end{equation}
Equivalently,
\begin{equation}
\langle O^2\rangle_\alpha^{\rm ren}
=
-\frac{2c_4}{\pi L}
\cot^3\alpha
+
O(\cot^4\alpha)
\simeq
-\frac{0.124}{L^2}
\cot^3\alpha.
\end{equation}
Numerical differentiation of \eqref{eq:Eren} gives
\begin{equation}
\frac{
2\partial_\lambda
\mathcal E_\alpha^{\rm ren}}
{\lambda^3}
=
-1.478,\quad
-1.524,\quad
-1.540
\end{equation}
at $\lambda L=5\times10^{-2},2\times10^{-2},10^{-2}$, respectively, approaching $-8c_4L$.

The cubic power has a simple origin. Expanding the determinant in the double-trace coupling, the first three terms are local ultraviolet contributions and are removed by the counterterms. The first finite contribution to the vacuum energy is therefore the quartic term $\lambda^4(\widehat G_N^E)^4$. Differentiating with respect to $\lambda$ leaves a leading $\lambda^3$ dependence. The exponent is thus fixed by the structure of the renormalized flow.
At the opposite end of the stable range, differentiating the square-root cusp \eqref{eq:cusp} gives
\begin{equation}
\langle O^2\rangle_\alpha^{\rm ren}
\propto
(\alpha_{\rm crit}-\alpha)^{-1/2}.
\label{eq:O2crit}
\end{equation}
The divergence reflects the growing fluctuations of the homogeneous mode as its gap closes. The same soft singularity will reappear in the Mellin analysis of Section \ref{sec:celestial}.

The ultraviolet behavior is very different. As the conformal boundary is approached, \cite{Morley:2020ayr}
\begin{equation}
\langle\phi^2\rangle_\alpha
\to
\frac{5}{48\pi^2L^2},
\qquad
\alpha\neq0.
\label{eq:UVuniv}
\end{equation}
This is the Neumann value. From the boundary viewpoint the result follows from the relevance of the deformation. At large frequency and angular momentum,
\begin{equation}
\widehat G_N^E
\sim
\frac{L}{
\sqrt{\Omega^2+(\ell+1)^2}},
\end{equation}
and the positive-mass-dimension coupling becomes negligible compared with the inverse propagator. The deformation modifies the infrared while leaving the leading short-distance coefficient unchanged.

\footnote{Before subtraction, expanding the coincident two-point function in $\lambda$ gives an $\alpha$-independent leading divergence. The first $\alpha$-dependent ultraviolet term is proportional to $-\lambda\sum_\ell(2\ell+1)\int\frac{d\Omega}{2\pi}(\widehat G_N^E)^2$. Its dependence on the boundary condition enters only through the explicit coupling. After the local terms are subtracted, the first finite contribution is the quartic term discussed above.}

The two limits therefore give a consistent picture. The leading short-distance structure is independent of $\alpha$, while the finite spectral response depends strongly on the boundary condition. The same separation is encoded in the resolvent $\widehat G_\alpha=\widehat G_N/(1+\lambda\widehat G_N)$ and in the position-space integral relation of \cite{Morley:2020ayr}.

The comparison with $T\bar T$ is useful only at the level of solvability. In both cases the deformed spectrum is determined from undeformed data. The mechanisms are otherwise different, since $T\bar T$ is irrelevant and reorganizes the ultraviolet spectrum \cite{Zamolodchikov:2004ce,Smirnov:2016lqw}, whereas the present deformation is relevant and leaves the ultraviolet fixed. Its exact solvability follows from Gaussianity rather than two-dimensional integrability. The Dirichlet endpoint again stands apart because $\cot\alpha$ diverges and the natural description changes to the standard quantization.

\subsection{Entropy, heat capacity and the soft-mode scaling region}
\label{sec:heatcap}

The entropy and heat capacity of the deformation, measured relative to Neumann, are
\begin{equation}
\Delta S(\alpha,T)
=
-\partial_TF_{\rm th},
\qquad
\Delta C(\alpha,T)
=
-T\partial_T^2F_{\rm th}.
\label{eq:SC}
\end{equation}
Both are ultraviolet finite because the counterterms are temperature independent.

Near the critical point the homogeneous mode has frequency
\begin{equation}
\omega_0
=
2B\sqrt{\alpha_{\rm crit}-\alpha},
\end{equation}
and its heat capacity is
\begin{equation}
\Delta C^{\rm soft}(\alpha,T)
=
\left(
\frac{\omega_0}{T}
\right)^2
\frac{
e^{\omega_0/T}}
{
\left(
e^{\omega_0/T}-1
\right)^2},
\label{eq:Csoft}
\end{equation}
where the ratio $\omega_0/T$ controls the crossover. For $T\ll\omega_0$,
\begin{equation}
\Delta C^{\rm soft}
\simeq
\left(
\frac{\omega_0}{T}
\right)^2
e^{-\omega_0/T},
\end{equation}
and the mode is thermally frozen out. For $\omega_0\ll T$,
\begin{equation}
\Delta C^{\rm soft}\to 1.
\end{equation}
Thus, at any fixed nonzero temperature, the soft mode approaches the classical equipartition value as $\alpha\to\alpha_{\rm crit}$.
The crossover occurs when $T\sim\omega_0$, or $T^2\sim 4B^2 (\alpha_{\rm crit}-\alpha)$. This defines the soft scaling region around the zero-temperature endpoint. After the appropriate overall powers of $T$ are factored out, the singular thermodynamic functions depend only on the ratio $\omega_0/T$, as expected for a Gaussian mode with no additional anomalous scale.

Equation \eqref{eq:Csoft} is only the soft contribution to the heat capacity. Modes with $\ell\geq1$ remain gapped at $\alpha_{\rm crit}$ and produce a background that is smooth in $\alpha$. At high temperature, differentiating \eqref{eq:highT} gives
\begin{equation}
\Delta C
\to
-\frac{2\pi}{3}
\cot\alpha\,LT,
\end{equation}
and therefore
\begin{equation}
\Delta C
\to
\frac{4}{3}LT
\qquad
\text{at }\alpha=\alpha_{\rm crit},
\qquad
T\gg L^{-1}.
\label{eq:Cbackground}
\end{equation}
The full heat capacity thus grows linearly at high temperature and does not approach unity. At the Neumann point,
\begin{equation}
\Delta C(\pi/2,T)=0
\end{equation}
identically. Near $\alpha_{\rm crit}$ the numerical values of the full heat capacity has an asymptotic slope $1.34$, close to the predicted $4/3$. The soft oscillator dominates the nonanalytic low-temperature part, while the higher modes supply the smooth background.

\begin{figure}[t]
\centering
\includegraphics[width=\textwidth]{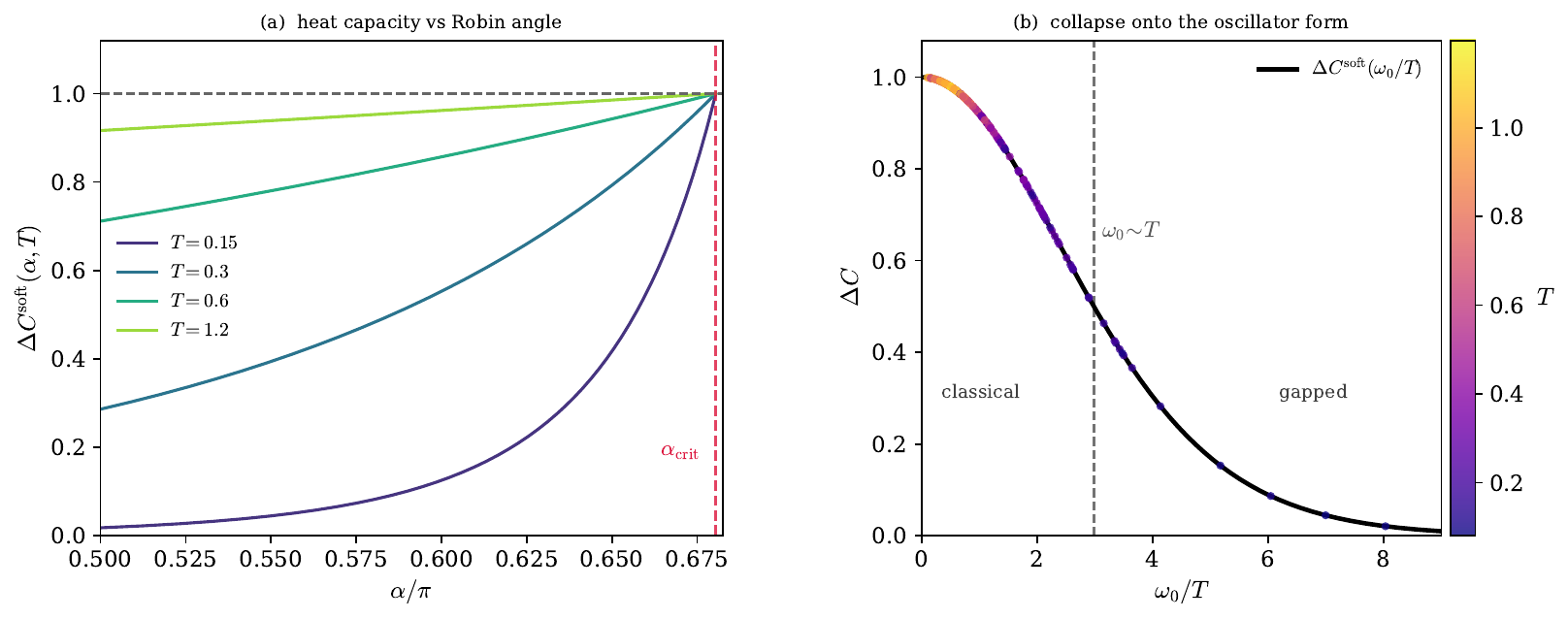}
\caption{Soft-mode contribution to the heat capacity near the critical angle ($L=1$). Both panels show $\Delta C^{\rm soft}$ of \eqref{eq:Csoft}, not the full heat capacity. \textbf{(a)} $\Delta C^{\rm soft}$ as a function of $\alpha$ for several temperatures. It approaches the classical value $1$ as $\alpha\to\alpha_{\rm crit}$ at every $T>0$ and is exponentially suppressed when $T\ll\omega_0$. \textbf{(b)} The same data plotted against $\omega_0/T$. The crossover at $\omega_0\sim T$ marks the soft scaling region, whose boundary behaves as $T\propto(\alpha_{\rm crit}-\alpha)^{1/2}$.}
\label{fig:heatcap}
\end{figure}

The exact results above rely on the absence of bulk self-interactions. With an interaction such as $\lambda_4\phi^4$, the dressed inverse propagator becomes
\begin{equation}
\widehat G_\alpha^{-1}
=
\widehat G_N^{-1}
+\lambda
+\Sigma(\omega,\ell),
\end{equation}
where the self-energy depends on the coupling and temperature through the interacting propagators. The flow is then no longer autonomous, and the determinant \eqref{eq:DeltaF} no longer gives the complete answer \cite{Gubser:2002zh,Diaz:2007an}.

The main question is whether the homogeneous static channel remains the first mode to soften once $\Sigma(0,\ell)$ is included. Two features of the free result need not survive unchanged. A temperature-dependent self-energy can move the instability threshold, so the vertical line in the free phase diagram will generically bend. Moreover, on a compact sphere a finite quartic interaction regulates the Gaussian zero-mode divergence. A sharp symmetry-breaking transition would then require an additional large-$N$, infinite-volume or semiclassical limit. At weak coupling it is nevertheless natural to expect the homogeneous channel to remain the leading soft direction unless the self-energy changes the ordering of the angular gaps. The free result provides the reference point for this interacting problem.

\section{The flat-space limit and a celestial perspective}
\label{sec:celestial}

The conformal boundary of global AdS has the same angular $S^2$ that labels null directions in the flat-space limit. Restoring $L$, define the physical frequency and angular momentum scale by
\begin{equation}
w=\frac{\Omega}{L},
\qquad
k=\frac{\ell+1}{L}.
\end{equation}
The AdS flat-space limit \cite{Penedones:2010ue,Hijano:2019qmi} takes $L\to\infty$ at fixed $(w,k)$. A null ray emitted from the centre reaches the boundary at global time $t=\pi/2$, so the scaling
\begin{equation}
t
=
\frac{\pi}{2}
+
\frac{u}{L}
\end{equation}
resolves retarded time near future null infinity. Boundary frequencies conjugate to $u$ then become the physical energies of the Carrollian description on $\mathcal I\simeq\mathbb R_u\times S^2$ \cite{Donnay:2022aba,Bagchi:2022emh,Alday:2024yyj,Lipstein:2025jpa}. A Mellin transform in the energy organizes this data by boost weight \cite{Pasterski:2016qvg,Strominger:2017zoo,Raclariu:2021zjz}.

At fixed $(w,k)$, both $\Omega$ and $\ell$ grow with $L$. Again using
\begin{equation}
\frac{\Gamma(z)}{\Gamma(z+1/2)}
\sim
z^{-1/2},
\end{equation}
the Neumann kernel becomes
\begin{equation}
\widehat G_N^E(\Omega,\ell)
\to
\frac{L}{
\sqrt{\Omega^2+(\ell+1)^2}}
=
\frac{1}{
\sqrt{w^2+k^2}}.
\label{eq:largeGN}
\end{equation}
The dressed kernel therefore approaches
\begin{equation}
\widehat G_\alpha^E(w,k)
\to
\frac{1}{
\sqrt{w^2+k^2}+\lambda},
\qquad
\lambda=\frac{\cot\alpha}{L}.
\label{eq:flatkernel}
\end{equation}
This is the momentum-space propagator of a three-dimensional dimension-one operator with its inverse shifted by the double-trace coupling.

If $\alpha$ is held fixed,
\begin{equation}
\lambda
=
\frac{\cot\alpha}{L}
\to0.
\end{equation}
Fixed-energy observables therefore approach the Neumann result. Expanding \eqref{eq:flatkernel},
\begin{equation}
\frac{
\delta\widehat G_\alpha}
{\widehat G_N}
\simeq
-\frac{\cot\alpha}{
L\sqrt{w^2+k^2}}
+
O(L^{-2}).
\label{eq:leadingcorr}
\end{equation}
The leading Robin correction is thus suppressed by one power of $1/L$, at the same order as the first finite-curvature corrections.

Three scalings should be distinguished. The fixed-$(w,k)$ limit requires $\Omega,\ell\sim L$ and probes the planar high-energy region of the global correlator. The soft sector instead keeps $\Omega$ and $\ell$ finite as $L\to\infty$, so that $w,k=O(L^{-1})$ and the global mode structure remains visible. Finally, one may keep the physical coupling $\lambda$ fixed, while in that case
\begin{equation}
\cot\alpha
\sim
\lambda L,
\end{equation}
so the Robin angle must scale with $L$ rather than remain fixed. These are distinct limits.
The fixed-energy flat-space statement should not be confused with a statement about the Mellin transform over all energies. A Mellin transform is sensitive to the soft endpoint of the energy integral and can retain information that disappears pointwise at fixed energy. We therefore use the Mellin transform below as a spectral diagnostic of the soft sector, rather than identifying it with a complete celestial correlator of the interacting conformal-primary basis. This distinction is important because the scalar considered here is free and has no nontrivial flat-space scattering amplitude.

In terms of the physical coupling, the negative-coupling stable window
\begin{equation}
\lambda
>
-\frac{2}{\pi L}
\end{equation}
shrinks to zero as $L\to\infty$. At finite $L$, the curvature of the sphere regulates the static homogeneous response,
\begin{equation}
\widehat G_N^E(0,0)
=
\frac{\pi L}{2}.
\end{equation}
The flat kernel \eqref{eq:flatkernel}, by contrast, becomes singular as the Euclidean momentum tends to zero. For $\lambda<0$, its denominator vanishes on
\begin{equation}
w_E^2+k^2
=
\lambda^2.
\end{equation}
After analytic continuation,
\begin{equation}
w_L^2
=
k^2-\lambda^2.
\end{equation}
Modes with $k<|\lambda|$ therefore have imaginary Lorentzian frequency, with growth rate bounded by $|\lambda|$.

\footnote{Replacing the exact spherical response by the flat kernel with $k=(\ell+1)/L$ would predict the first instability from $1+\lambda L=0$, or $\cot\alpha=-1$. The exact homogeneous response is $\widehat G_N^E(0,0)=\pi L/2$, so the true threshold is instead $\cot\alpha=-2/\pi$.}

The vacuum energy has the same infrared character. Restoring dimensions,
\begin{equation}
\mathcal E_\alpha^{\rm ren}
\simeq
-\frac{0.195}{L}
\cot^4\alpha
\end{equation}
near Neumann. This contribution is $O(L^{-1})$, comparable to the energy of a single soft quantum, and is not extensive in the boundary area. Its energy density scales as
\begin{equation}
\frac{
\mathcal E_\alpha^{\rm ren}}
{4\pi L^2}
=
O(L^{-3}).
\end{equation}
At the cusp the singular contribution is
\begin{equation}
E_{\rm soft}
=
\frac{\omega_0}{2},
\end{equation}
with the dimensionful gap
\begin{equation}
\omega_0
=
\frac{2B}{L}
\sqrt{\alpha_{\rm crit}-\alpha}.
\end{equation}

The soft thermodynamics has a nontrivial flat-limit scaling if
\begin{equation}
\vartheta
\equiv
LT
\end{equation}
is kept fixed. Then
\begin{equation}
\frac{\omega_0}{T}
=
\frac{
2B\sqrt{\alpha_{\rm crit}-\alpha}}
{\vartheta}
\end{equation}
remains finite. The soft free energy and heat capacity therefore retain finite scaling functions of $(\alpha,\vartheta)$. At fixed physical $T>0$, on the other hand, $\omega_0/T\to0$, so the soft mode is driven into its classical regime and
\begin{equation}
\Delta C^{\rm soft}\to 1.
\end{equation}

From the Carrollian viewpoint the critical excitation is the static homogeneous component of the boundary field. It is independent of retarded time $u$ and lies in the $\ell=0$ sector. Its characteristic frequency is $O(L^{-1})$, so it is pushed to the zero-energy corner of the flat-space data. This is the same soft-frequency region in which conformally soft operators and asymptotic symmetries are naturally organized \cite{Donnay:2018neh,Pasterski:2021rjz,Hijano:2020szl}.
We now characterize this compression in Mellin space. Let $O_0(t)$ denote the $\ell=0$ projection of the boundary operator and define, in units $L=1$,
\begin{equation}
M_\alpha(\Delta)
\equiv
\int_0^\infty
d\Omega\,
\Omega^{\Delta-1}
\widehat G_\alpha^E(\Omega,0),
\qquad
0<\operatorname{Re}\Delta<1.
\label{eq:mellindef}
\end{equation}
This is the Mellin transform of one angular-momentum channel. It is not, by itself, a full celestial correlator in the conformal-primary basis \cite{Pasterski:2017kqt}. For the free scalar, such a celestial correlator remains distributional in the boost weights.
For the flat kernel and $\lambda>0$,
\begin{equation}
M_\alpha^{\rm flat}(\Delta)
=
\int_0^\infty
d\Omega\,
\frac{\Omega^{\Delta-1}}{\Omega+\lambda}
=
\frac{
\pi\lambda^{\Delta-1}}
{\sin\pi\Delta}.
\label{eq:mellinflat}
\end{equation}
The expression defines a meromorphic continuation in $\Delta$ and may itself be continued in $\lambda$ once a branch and, for negative $\lambda$, a pole prescription are chosen.

The resolvent expansion
\begin{equation}
\widehat G_\alpha
=
\sum_{m\geq0}
(-\lambda)^m
\widehat G_N^{\,m+1}
\end{equation}
makes the ultraviolet pole structure transparent. Since $\widehat G_N(\Omega,0)\sim\Omega^{-1}$, the $m$-th term behaves as
\begin{equation}
(-\lambda)^m
\Omega^{-m-1}.
\end{equation}
Its Mellin transform has a pole at $\Delta=m+1$, with
\begin{equation}
\operatorname{Res}_{\Delta=n}
M_\alpha^{\rm flat}
=
(-1)^n
\lambda^{n-1}.
\end{equation}
Within this positive-integer ultraviolet sequence,
\begin{equation}
\operatorname{Res}_{\Delta=1}
M_\alpha^{\rm flat}
=
-1
\end{equation}
is universal, and the Robin dependence first appears at $\Delta=2$.

The infrared pole is different. In the $L=1$ convention,
\begin{equation}
\operatorname{Res}_{\Delta=0}
M_\alpha^{\rm flat}
=
\frac{1}{\lambda}
=
\tan\alpha,
\end{equation}
which is the zero-momentum static response of the flat channel. For positive $\lambda$, along the principal series $\Delta=1+\ii\nu$, the magnitude of $\lambda^{\ii\nu}$ is unity, so the coupling enters \eqref{eq:mellinflat} only through a phase. For negative $\lambda$, the result depends on the branch and pole prescription and develops an imaginary part associated with the unstable flat-space mode.

At finite $L$, the homogeneous Neumann kernel is
\begin{equation}
\widehat G_N^E(\Omega,0)
=
\frac{\tanh(\pi\Omega/2)}{\Omega}
=
\frac{1}{\Omega}
+
O(e^{-\pi\Omega})
\label{eq:finiteL_homo}
\end{equation}
in $L=1$ units. The difference from the flat asymptotic form is exponentially small rather than a power series in $1/\Omega$. Consequently, in these dimensionless variables,
\begin{equation}
\operatorname{Res}_{\Delta=n}
M_\alpha
=
(-1)^n
(\cot\alpha)^{n-1}
\label{eq:finiteL_res}
\end{equation}
for the positive-integer ultraviolet poles. Finite curvature changes the infrared structure but introduces no additional power corrections into these ultraviolet residues.

At small frequency,
\begin{equation}
\widehat G_N^E(\Omega,0)
=
\frac{\pi}{2}
-
\frac{\pi^3}{24}\Omega^2
+
\frac{\pi^5}{240}\Omega^4
+\cdots.
\label{eq:GEN_IR}
\end{equation}
Define
\begin{equation}
\varepsilon
=
1+\frac{\pi}{2}\lambda
\end{equation}
in $L=1$ units. Expanding the dressed kernel gives
\begin{equation}
\operatorname{Res}_{\Delta=0}
M_\alpha
=
\frac{\pi/2}{\varepsilon}
=
\chi_O,
\end{equation}
\begin{equation}
\operatorname{Res}_{\Delta=-2}
M_\alpha
=
-\frac{\pi^3/24}{\varepsilon^2},
\end{equation}
and
\begin{equation}
\operatorname{Res}_{\Delta=-4}
M_\alpha
=
\frac{\pi^5/240}{\varepsilon^2}
-
\frac{
\lambda(\pi^3/24)^2}
{\varepsilon^3}.
\label{eq:softresidues}
\end{equation}
More generally,
\begin{equation}
\operatorname{Res}_{\Delta=-2m}
M_\alpha
=
O\left(
\varepsilon^{-(m+1)}
\right)
\end{equation}
at leading order near the critical point, and successive negative even weights therefore probe increasingly high powers of the diverging static response.

At the Neumann point,
\begin{equation}
M_{\pi/2}(\Delta)
=
-2\pi^{1-\Delta}
\left(
1-2^{2-\Delta}
\right)
\Gamma(\Delta-1)
\zeta(\Delta-1).
\label{eq:mellinN}
\end{equation}
The trivial zeros of the zeta function cancel the gamma-function poles at the odd negative integers, leaving the infrared sequence
$\Delta=0,-2,-4,\ldots$, and the residues at $\Delta=1$ and $\Delta=0$ agree with the ultraviolet and infrared limits above.

The higher angular-momentum channels have the same meromorphic character. The recursion relation \eqref{eq:recursion} expresses each Neumann kernel as a rational function of $\Omega^2$ multiplying either $\tanh(\pi\Omega/2)$ or $\coth(\pi\Omega/2)$. The homogeneous channel is distinguished only because its infrared response becomes singular at $\alpha_{\rm crit}$.
Near the critical angle, the soft-pole approximation gives a universal Mellin scaling form. Using
\begin{equation}
\int_0^\infty
d\Omega\,
\frac{\Omega^{\Delta-1}}
{\Omega^2+\omega_0^2}
=
\frac{\pi}{2}
\frac{
\omega_0^{\Delta-2}}
{\sin(\pi\Delta/2)},
\end{equation}
one obtains
\begin{equation}
M_\alpha^{\rm soft}(\Delta)
\simeq
\frac{\pi^2}{4\kappa}
\frac{
\omega_0^{\Delta-2}}
{\sin(\pi\Delta/2)}
=
\frac{
3\omega_0^{\Delta-2}}
{\sin(\pi\Delta/2)}
\propto
(\alpha_{\rm crit}-\alpha)^{(\Delta-2)/2},
\label{eq:mellincrit}
\end{equation}
where $\kappa=\pi^2/12$. The exponent is linear in the Mellin weight because $\omega_0$ is the only scale in the singular sector.

At $\Delta=0$, \eqref{eq:mellincrit} should be read through its residue. It gives
\begin{equation}
\operatorname{Res}_{\Delta=0}
M_\alpha^{\rm soft}
\propto
\omega_0^{-2}
\propto
(\alpha_{\rm crit}-\alpha)^{-1},
\end{equation}
which reproduces \eqref{eq:chi0}. At $\Delta=1$, after subtracting the universal ultraviolet pole of the full Mellin transform, the critical part scales as
\begin{equation}
M_\alpha^{\rm sing}(1)
\propto
\omega_0^{-1}
\propto
(\alpha_{\rm crit}-\alpha)^{-1/2},
\end{equation}
which is the same inverse-square-root singularity obtained by differentiating the Casimir cusp.

The slope
\begin{equation}
\frac{d}{d\Delta}
\frac{\Delta-2}{2}
=
\frac12
\end{equation}
is the gap exponent. Expanding \eqref{eq:mellincrit} about $\Delta=-2m$ gives
\begin{equation}
\operatorname{Res}_{\Delta=-2m}
M_\alpha
\simeq
\frac{6}{\pi}
(-1)^m
\left(
\frac{\kappa}{\varepsilon}
\right)^{m+1}.
\label{eq:m2mresidue}
\end{equation}
For example,
\begin{equation}
\operatorname{Res}_{\Delta=-2}
M_\alpha
\simeq
-\frac{\pi^3}{24\varepsilon^2},
\qquad
\operatorname{Res}_{\Delta=-4}
M_\alpha
\simeq
\frac{\pi^5}{288\varepsilon^3}.
\label{eq:resexample}
\end{equation}
The near-critical Mellin profile therefore generates the leading singular behavior of the full tower of soft residues.

The same expression may be analytically continued into the unstable regime. For $\alpha>\alpha_{\rm crit}$ the Euclidean kernel develops a pole at
\begin{equation}
\Omega_*
=
2B
\sqrt{\alpha-\alpha_{\rm crit}},
\end{equation}
corresponding to the Lorentzian growth rate of the unstable homogeneous mode. Continuing $\omega_0\to\ii\Omega_*$ and choosing a pole prescription gives
\begin{equation}
\operatorname{Im}
M_\alpha^{\rm soft}(\Delta)
=
\mp
\frac{\pi^2}{4\kappa}
\Omega_*^{\Delta-2}.
\label{eq:immellin}
\end{equation}
The stable-side Mellin profile therefore measures the closing gap, while its unstable-side continuation measures the growth rate of the negative mode.

The soft-pole approximation rapidly becomes accurate near the endpoint. At $\Delta=1/2$, the ratio of the exact Mellin transform to \eqref{eq:mellincrit} differs from unity by $1.2\times10^{-2}$ at $\varepsilon=10^{-2}$ and by $1.7\times10^{-5}$ at $\varepsilon=10^{-6}$. At finite temperature the same mode appears as the static Matsubara contribution and produces the coefficient $T/2$ in the logarithmic thermal singularity.

There is also a useful flat-space interpretation of the homogeneous mode. As $L\to\infty$, the conformal mass $-2/L^2$ vanishes and the limiting free massless scalar has the shift symmetry
\begin{equation}
\phi\rightarrow\phi+c.
\end{equation}
Its zero-frequency sector is related to the asymptotic charges and scalar memory observables discussed in \cite{Campiglia:2017dpg,Campiglia:2017xkp,Henneaux:2018cst}. At finite $L$ the corresponding homogeneous mode is lifted by the AdS curvature and Robin boundary condition, and its effective quadratic coefficient is proportional to
\begin{equation}
1+\lambda
\widehat G_N^E(0,0),
\end{equation}
while $\alpha_{\rm crit}$ this coefficient changes sign.

The analogy with scalar memory is limited, since in the unstable free theory the homogeneous field grows rather than settling to a finite late-time displacement \cite{Hamada:2017atr}. A stabilizing bulk interaction could instead generate a finite minimum. In a mean-field treatment one would then expect
\begin{equation}
\langle O\rangle
\propto
(\alpha-\alpha_{\rm crit})^{1/2}.
\end{equation}
Global AdS with Robin boundary conditions can therefore be viewed as an infrared regularization of this soft scalar sector. The continuous low-frequency region is replaced by modes separated by $O(L^{-1})$, while the Robin parameter tunes the lowest mode through zero. No Goldstone theorem is involved in the free problem, and the relevant analogy is simply the emergence of a soft zero-frequency direction.

These statements have a limited celestial scope for several reasons. The free scalar has no nontrivial flat-space $S$-matrix and hence no nontrivial celestial scattering amplitudes on which the transition could act. The hyperbolic-slicing constructions of \cite{deBoer:2003vf,Casali:2022fro} are also conceptually different: they embed AdS$_3$ slices in Minkowski space, whereas the present discussion takes the flat-space limit of AdS$_4$.
The double-trace interpretation of mixed boundary conditions in celestial and wedge settings \cite{Fukada:2023vjt} is compatible with the present limit, but here the deformation is first resummed exactly at finite $L$. The main conclusion is that fixed-energy two-point data lose their Robin dependence, while the soft-frequency sector retains it. In Mellin space the critical behavior is encoded in poles at the soft weights through \eqref{eq:mellincrit}.

However, the gravitational analogue is correspondingly well posed. Relaxing the standard reflective boundary conditions of AdS$_4$ gravity leads to the leaky $\Lambda$-BMS family, whose asymptotic symmetry algebra contracts to BMS$_4$ in the flat limit \cite{Compere:2019bua,Compere:2020lrt}. The scalar problem provides a tractable example in which the effect of relaxed boundary conditions on the spectrum and thermodynamics can be followed explicitly. It is natural to ask whether analogous graviton boundary conditions possess a parameter at which a distinguished soft mode reaches zero frequency at finite AdS radius. Equation \eqref{eq:mellincrit} provides a scalar benchmark for such a calculation.

\section{Conclusions and outlook}
\label{sec:conclusions}

A conformally coupled scalar in global AdS$_4$ with Robin boundary conditions is governed by a single response function in each $(\omega,\ell)$ channel. Because the conformal map to the half-Einstein static universe is exact, the boundary integral equation of \cite{Deutsch:1978sc} diagonalizes on $\mathbb{R}\times S^2$. The resulting boundary propagator,
\begin{equation}
\widehat{\mathcal G}_\alpha
=
\frac{\widehat{\mathcal G}_N}
{1+\lambda\widehat{\mathcal G}_N},
\qquad
\lambda=\frac{\cot\alpha}{L},
\end{equation}
is nonperturbative in the Robin parameter. Its poles give the normal frequencies, its residues reproduce the mode expansion, its static denominator fixes the stability endpoint, and its determinant determines the Gaussian free energy.

The free bulk theory makes the double-trace flow exactly Gaussian. The determinant \eqref{eq:DeltaF} therefore receives no further corrections within this sector. Its temperature-independent Casimir part contains three ultraviolet divergences, removed by local boundary counterterms. The first scheme-independent coefficient occurs at quartic order and is
\begin{equation}
c_4
\simeq
\frac{0.195}{L}.
\end{equation}
The renormalized vacuum energy approaches the critical angle with a finite square-root cusp. Its coefficient,
\begin{equation}
B\simeq0.819,
\end{equation}
is predicted by the soft expansion and agrees with the numerical fit.

For every finite Robin coupling, the leading $T^4$ and $T^3$ terms in the absolute high-temperature free energy are independent of $\alpha$ and cancel in $F_\alpha-F_N$. The first surviving Robin-dependent contribution is
\begin{equation}
F_{\rm th}
\sim
\frac{\pi}{3}
\cot\alpha\,LT^2.
\end{equation}
The Dirichlet endpoint is a nonuniform strong-coupling limit and instead has a leading $T^3$ difference from Neumann. Differentiation with respect to $\lambda$ gives the renormalized double-trace expectation value. Its cubic behavior near Neumann follows from the fact that the first three terms in the coupling expansion are local ultraviolet counterterms and the first finite term is quartic.

A single mode controls the stability endpoint. The quantity $1+\lambda \widehat G_N^E(0,\ell)$ is proportional to the inverse static susceptibility, and the homogeneous channel has the largest static response. Its gap closes at $\alpha_{\rm crit} =\pi-\arctan(\pi/2)$, reproducing the classical threshold of \cite{Morley:2020ayr}. The frequency behaves as
\begin{equation}
\omega_0
\propto
(\alpha_{\rm crit}-\alpha)^{1/2},
\end{equation}
while
\begin{equation}
\chi_0
\propto
(\alpha_{\rm crit}-\alpha)^{-1}.
\end{equation}
The Gaussian gap and susceptibility exponents are therefore $1/2$ and $1$, respectively.
The singular thermodynamics is the free energy of this one oscillator. Its zero-point energy gives the $T=0$ cusp, its classical thermal free energy gives the logarithmic divergence at any $T>0$, and its heat capacity is controlled by the ratio $\omega_0/T$. The nonanalyticity is consequently a single-mode effect rather than a collective rearrangement of the spectrum. A Hagedorn interpretation is not appropriate.

The ultraviolet and infrared behaviors provide complementary checks. The renormalized vacuum polarization approaches the same boundary value for every non-Dirichlet Robin condition \cite{Morley:2020ayr}, as expected for a relevant deformation that leaves the leading short-distance structure unchanged. The dependence on $\alpha$ instead appears in the low-lying normal modes and the finite part of the double-trace response.

The flat-space limit sharpens this distinction. At fixed physical energy,
\begin{equation}
\lambda
=
\frac{\cot\alpha}{L}
\to0,
\end{equation}
so the leading two-point function becomes independent of the Robin angle. The deformation survives only in the soft-frequency region $w=O(L^{-1})$. A Mellin transform of the homogeneous channel provides a useful spectral description of this region. Its ultraviolet residues are fixed by the large-frequency expansion, while its infrared residues diverge as the critical point is approached. Near the endpoint they are organized by
\begin{equation}
M_\alpha^{\rm soft}(\Delta)
\propto
(\alpha_{\rm crit}-\alpha)^{(\Delta-2)/2}.
\end{equation}
For $\alpha>\alpha_{\rm crit}$ the analytic continuation of the same expression encodes the growth rate of the unstable mode.
\
This Mellin transform should not be identified with a full celestial scattering correlator. The scalar is free and has no nontrivial flat-space $S$-matrix. Its role here is to organize the soft-frequency two-point data and to show how the single Gaussian scale $\omega_0$ appears across different Mellin weights.

Two assumptions suggest natural extensions. The first is the absence of bulk self-interactions. Interactions generate a self-energy,
\begin{equation}
\widehat G_\alpha^{-1}
=
\widehat G_N^{-1}
+\lambda
+\Sigma,
\end{equation}
so the flow is no longer autonomous and the determinant no longer captures the complete quantum theory. A temperature-dependent self-energy will generically move the instability threshold, while a finite quartic interaction on the compact sphere regulates the Gaussian zero-mode divergence. The main question is whether the static homogeneous channel remains the first mode to soften. At weak coupling this is the natural continuation of the free result, but it need not hold if interactions reorder the angular gaps. A stabilized theory could also support a nonzero order parameter beyond the Gaussian endpoint.
The second assumption is conformal coupling. The mixed family exists throughout the Breitenlohner--Freedman window, but the direct identification of the two asymptotic coefficients with the boundary value and normal derivative of a smooth rescaled field uses
\begin{equation}
\Delta_+-\Delta_-=1.
\end{equation}
For general mass the boundary condition is most naturally written in terms of the asymptotic coefficients,
\begin{equation}
B+g_\Delta A=j,
\qquad
g_\Delta
=
\lambda_\Delta
L^{\Delta_+-\Delta_-},
\end{equation}
rather than as an elementary local Robin condition on the rescaled field. The radial solutions remain expressible in hypergeometric form, but the conformal map and simple image representation used here no longer provide the same simplifications. The resolvent construction itself should persist because it follows from the quadratic boundary deformation rather than from conformal coupling.
The same ultraviolet--infrared separation should also organize the renormalized stress-energy tensor studied in \cite{Morley:2023mmy}. Its thermal component would give direct access to the pressure and provide a further test of the equation of state.

Finally, it is natural to ask whether an analogous softening mechanism occurs in gravity. Relaxing reflective AdS$_4$ boundary conditions leads to the leaky $\Lambda$-BMS family. The scalar theory studied here gives a benchmark in which relaxed boundary conditions, spectral stability and soft thermodynamics can all be computed explicitly. The corresponding gravitational question is whether a distinguished graviton mode can be tuned to zero frequency at finite AdS radius and, if so, how that softening is encoded in the flat-space limit.

\appendix

\section{The Euclidean Neumann response from the regular solution}
\label{app:propagator}

Equation \eqref{eq:GEN} is used throughout, so we record its derivation. Dividing the spectral condition $D_\alpha(\omega,\ell)=0$ by $\sin\alpha$ gives
\begin{equation}
1
+
\cot\alpha\,
\frac{
\psi_{\omega\ell}(\pi/2)}
{\psi'_{\omega\ell}(\pi/2)}
=
0.
\end{equation}
Comparison with \eqref{eq:resolvent_pole_condition}, together with $\lambda=\cot\alpha/L$, gives
\begin{equation}
\widehat{\mathcal G}_N(\omega,\ell)
=
L
\frac{
\psi_{\omega\ell}(\pi/2)}
{\psi'_{\omega\ell}(\pi/2)}.
\label{eq:app_ratio}
\end{equation}
This ratio is independent of the normalization of the regular solution and has the required dimension $L$.

In Euclidean signature the regular solution is \eqref{eq:regular_euclidean}. Writing $x=\cos\rho$, so that
\begin{equation}
\partial_\rho
=
-\frac{d}{d x}
\qquad
\text{at }\rho=\frac{\pi}{2},
\end{equation}
the prefactor $(1-x^2)^{-1/4}$ equals unity at $x=0$ and has vanishing derivative there. Thus
\begin{equation}
p_{\Omega\ell}\left(\frac{\pi}{2}\right)
=
P^{-\ell-1/2}_{\ii\Omega-1/2}(0),
\end{equation}
and
\begin{equation}
\left.
\partial_\rho p_{\Omega\ell}
\right|_{\pi/2}
=
-
\left.
\frac{d}{d x}
P^{-\ell-1/2}_{\ii\Omega-1/2}(x)
\right|_{x=0}.
\end{equation}
The required values are
\begin{equation}
P^{-\mu}_{\nu}(0)
=
\frac{\sqrt{\pi}}
{
2^\mu
\Gamma\!\left(
\frac{\mu+\nu}{2}+1
\right)
\Gamma\!\left(
\frac{\mu-\nu}{2}+\frac12
\right)},
\end{equation}
and
\begin{equation}
\left.
\frac{d P^{-\mu}_\nu}{d x}
\right|_{x=0}
=
-
\frac{\sqrt{\pi}}
{
2^{\mu-1}
\Gamma\!\left(
\frac{\mu+\nu}{2}+\frac12
\right)
\Gamma\!\left(
\frac{\mu-\nu}{2}
\right)}.
\end{equation}
Setting
\begin{equation}
\mu=\ell+\frac12,
\qquad
\nu=\ii\Omega-\frac12,
\end{equation}
the gamma functions pair into complex conjugates for real $\Omega$, giving
\begin{equation}
p_{\Omega\ell}\left(\frac{\pi}{2}\right)
=
\frac{\sqrt{\pi}}{2^{\ell+1/2}}
\left|
\Gamma\!\left(
\frac{\ell+2+\ii\Omega}{2}
\right)
\right|^{-2},
\end{equation}
and
\begin{equation}
\left.
\partial_\rho p_{\Omega\ell}
\right|_{\pi/2}
=
\frac{\sqrt{\pi}}{2^{\ell-1/2}}
\left|
\Gamma\!\left(
\frac{\ell+1+\ii\Omega}{2}
\right)
\right|^{-2}.
\end{equation}
Their ratio gives \eqref{eq:GEN}. Positivity for real $\Omega$ is manifest, and \eqref{eq:recursion} follows directly from $\Gamma(z+1)=z\Gamma(z)$. The elementary expressions \eqref{eq:GEN01} and static values \eqref{eq:static} follow from the reflection and duplication identities.

We also evaluated \eqref{eq:app_ratio} directly using Legendre functions of complex degree. The numerical result agrees with \eqref{eq:GEN} at the accuracy set by the finite-difference derivative. The agreement of the resulting zero-mode threshold with \cite{Morley:2020ayr} provides an independent check of the sign and normalization conventions.

\section{Power counting and the subtractions}
\label{app:powercount}

The ultraviolet subtractions in \eqref{eq:Eren} follow from the large-argument expansion of \eqref{eq:GEN}. Let
\begin{equation}
z
=
\frac{\ell+1+\ii\Omega}{2}.
\end{equation}
Then
\begin{equation}
\widehat G_N^E
=
\frac{L}{2}
\left|
\frac{\Gamma(z)}
{\Gamma(z+1/2)}
\right|^2,
\end{equation}
and
\begin{equation}
\frac{\Gamma(z)}
{\Gamma(z+1/2)}
=
z^{-1/2}
\left[
1+\frac{1}{8z}
+O(z^{-2})
\right].
\end{equation}
In units $L=1$,
\begin{equation}
\widehat G_N^E(\Omega,\ell)
=
\frac{1}{\nu}
+
\frac{k}{2\nu^3}
+
O(\nu^{-3}),
\qquad
k=\ell+1,
\qquad
\nu=\sqrt{\Omega^2+k^2}.
\label{eq:app_largenu}
\end{equation}

For a function depending only on $\nu$, the asymptotic mode sum becomes
\begin{equation}
\sum_\ell(2\ell+1)
\int_{-\infty}^{\infty}
\frac{d\Omega}{2\pi}
f(\nu)
\to
\frac{2}{\pi}
\int_0^\infty
d\nu\,
\nu^2f(\nu).
\label{eq:app_measure}
\end{equation}
Using the leading term of \eqref{eq:app_largenu} in the logarithm gives
\begin{equation}
\mathcal E_\alpha
\sim
\frac{1}{\pi}
\sum_{p\geq1}
\frac{(-1)^{p+1}\lambda^p}{p}
\int^\Lambda
d\nu\,
\nu^{2-p}.
\label{eq:app_orders}
\end{equation}
The terms with $p=1$ and $p=2$ diverge quadratically and linearly, respectively, while $p=3$ gives a logarithmic divergence. Every $p\geq4$ term converges.

The leading divergence is
\begin{equation}
\mathcal E_\alpha
\supset
\frac{\lambda\Lambda^2}{2\pi}.
\end{equation}
At $\lambda L=0.3$, a sharp cutoff gives
\begin{equation}
\frac{\mathcal E_\alpha}{\Lambda^2}
=
0.04748,\quad
0.04761,\quad
0.04768
\end{equation}
for $L\Lambda=200,400,800$, and these values approach $\lambda/(2\pi)= 0.047746$ with the expected inverse-cutoff corrections.

The divergences are local polynomials in $\lambda$ of degree one, two and three. They may be removed by counterterms of the form
\begin{equation}
\int d^3x\sqrt{\gamma}
\left(
a_1\lambda
+
a_2\lambda^2
+
a_3\lambda^3
\right),
\end{equation}
where
\begin{equation}
[a_1]=L^{-2},
\qquad
[a_2]=L^{-1},
\qquad
[a_3]=1.
\end{equation}
Their finite parts produce the residual scheme freedom
\begin{equation}
c_1\cot\alpha
+
c_2\cot^2\alpha
+
c_3\cot^3\alpha,
\end{equation}
and the subtraction prescription \eqref{eq:Eren} sets these finite terms to zero.

Two consequences are immediate. First, the subtracted expansion begins at $u^4$. By \eqref{eq:app_measure}, its radial integrand falls as $\nu^{-2}$, so the integral converges and $c_4$ is unambiguous. Second, because all subtractions are temperature independent, the thermal free energy, entropy and heat capacity are ultraviolet finite without further counterterms. The same reasoning makes every coefficient beyond cubic order in the expansion about the Neumann point scheme independent.

\section{Numerical evaluation}
\label{app:numerics}

All numerical quantities are constructed from \eqref{eq:GEN}. The gamma-function ratio is evaluated through $\log\Gamma$ to maintain stability at large $\ell$ and $\Omega$. Frequency integrals use adaptive quadrature with absolute and relative tolerances $10^{-11}$. Angular-momentum sums are completed with analytic large-$\ell$ tails.

For $c_4$, define
\begin{equation}
t(\ell)
=
\frac{2\ell+1}{8}
\int
\frac{d\Omega}{2\pi}
\left(
\widehat G_N^E
\right)^4
\end{equation}
in units $L=1$. Equation \eqref{eq:app_largenu} gives
\begin{equation}
t(\ell)
\sim
\frac{2\ell+1}
{32(\ell+1)^3}
\end{equation}
at large $\ell$. Summing through $\ell=1200$ and adding this tail gives $c_4L =0.19469$, of which the homogeneous channel contributes $0.1394$. The same value follows from the $\cot\alpha\to0$ limit of $\mathcal E_\alpha^{\rm ren}/\cot^4\alpha$ evaluated from the unexpanded \eqref{eq:Eren}. For the full renormalized vacuum energy, $\ell\leq400$ is sufficient once the subtractions have reduced the integrand to $O(u^4)$.

The thermal free energy is evaluated channel by channel as the Matsubara sum minus the zero-temperature integral. For each $\ell$, the frequency sum is truncated at $n\leq N$ with
\begin{equation}
2\pi NT
\gtrsim
3(\ell+2).
\end{equation}
The continuum integral is truncated at the matching midpoint frequency
\begin{equation}
2\pi T
\left(
N+\frac12
\right),
\end{equation}
so that the leading truncation errors cancel between the two terms. The angular-momentum sum is continued to $\ell_{\max}
\simeq 14LT+30$, beyond which the contribution is suppressed as $e^{-(\ell+1)/(LT)}$ when the dimensionful temperature is restored.

The high-temperature law \eqref{eq:highT} was checked at $\lambda L=-10^{-2}$. The values of $F_{\rm th}/(\lambda L^2T^2)$ quoted in Section \ref{sec:phase} approach $\pi/3$ with the expected inverse-temperature corrections. The double-trace expectation value was also checked by central differences of \eqref{eq:Eren} in $\lambda$, using a relative step $10^{-2}$.

Finally, the Legendre-function evaluation in Appendix \ref{app:propagator} was performed at thirty-digit precision. A one-sided boundary derivative with step $10^{-8}$ agrees with \eqref{eq:GEN} at the corresponding $10^{-8}$ level. The numerical code used for the figures reproduces all values quoted in the text.

\acknowledgments
The authors gratefully acknowledge support from the Graduate School of Brown University. We thank Yiru Wang for valuable discussions.

\bibliographystyle{jhep}
\bibliography{citation}
\end{document}